\documentclass[twocolumn]{aastex63}

\usepackage{amsmath}
\usepackage{graphicx}
\usepackage{xcolor}
\usepackage{gensymb}
\usepackage[caption=false]{subfig}
\usepackage{xcolor}
\usepackage{scrextend}
\usepackage{CJKutf8}

\newcommand{\beq}{\begin{equation}}
\newcommand{\eeq}{\end{equation}}
\newcommand{\bea}{\begin{eqnarray}}
\newcommand{\eea}{\end{eqnarray}}

\shortauthors{Q. Liu et al.}
\shorttitle{Wide-angle PSF}
\graphicspath{{./}{figures/}}

\begin{document}

\title{A Method To Characterize the Wide-Angle Point Spread Function of Astronomical Images}

\correspondingauthor{Qing Liu}
\email{qliu@astro.utoronto.ca}

\author[0000-0002-7490-5991]{Qing Liu \begin{CJK}{UTF8}{gbsn}(刘青)\end{CJK}}
\affil{David A. Dunlap Department of Astronomy \& Astrophysics, University of Toronto, 50 St. George St., Toronto, ON M5S 3H4, Canada}
\affil{Dunlap Institute for Astronomy and Astrophysics, University of Toronto, Toronto ON, M5S 3H4, Canada}

\author[0000-0002-4542-921X]{Roberto Abraham}
\affil{David A. Dunlap Department of Astronomy \& Astrophysics, University of Toronto, 50 St. George St., Toronto, ON M5S 3H4, Canada}
\affil{Dunlap Institute for Astronomy and Astrophysics, University of Toronto, Toronto ON, M5S 3H4, Canada}

\author[0000-0002-8931-4684]{Colleen Gilhuly}
\affil{David A. Dunlap Department of Astronomy \& Astrophysics, University of Toronto, 50 St. George St., Toronto, ON M5S 3H4, Canada}
\affil{Dunlap Institute for Astronomy and Astrophysics, University of Toronto, Toronto ON, M5S 3H4, Canada}

\author[0000-0002-8282-9888]{Pieter van Dokkum}
\affil{Department of Astronomy, Yale University, New Haven, CT 06520, USA}

\author[0000-0002-5236-3896]{Peter G. Martin}
\affil{Canadian Institute for Theoretical Astrophysics, University of Toronto, 60 St. George St., Toronto,
ON M5S 3H8, Canada}

\author[0000-0001-9592-4190]{Jiaxuan Li \begin{CJK}{UTF8}{gbsn}(李嘉轩)\end{CJK}}
\affil{Kavli Institute for Astronomy and Astrophysics, Peking University, 5 Yiheyuan Road, Haidian District, Beijing 100871, China}
\affil{Department of Astrophysical Sciences, Princeton University, 4 Ivy Lane, Princeton, NJ 08544, USA}

\author[0000-0003-4970-2874]{Johnny P. Greco}
\altaffiliation{NSF Astronomy \& Astrophysics Postdoctoral Fellow}
\affil{Center for Cosmology and AstroParticle Physics (CCAPP), The Ohio State University, Columbus, OH 43210, USA}

\author[0000-0002-2406-7344]{Deborah Lokhorst}
\affil{David A. Dunlap Department of Astronomy \& Astrophysics, University of Toronto, 50 St. George St., Toronto, ON M5S 3H4, Canada}
\affil{Dunlap Institute for Astronomy and Astrophysics, University of Toronto, Toronto ON, M5S 3H4, Canada}

\author[0000-0002-4175-3047]{Seery Chen}
\affil{David A. Dunlap Department of Astronomy \& Astrophysics, University of Toronto, 50 St. George St., Toronto, ON M5S 3H4, Canada}
\affil{Dunlap Institute for Astronomy and Astrophysics, University of Toronto, Toronto ON, M5S 3H4, Canada}

\author[0000-0002-1841-2252]{Shany Danieli}
\altaffiliation{NASA Hubble Fellow}
\affil{Department of Physics, Yale University, New Haven, CT 06520, USA}
\affil{Department of Astronomy, Yale University, New Haven, CT 06520, USA}
\affil{Yale Center for Astronomy and Astrophysics, Yale University, New Haven, CT 06511, USA}
\affil{Institute for Advanced Study, 1 Einstein Drive, Princeton, NJ 08540, USA}

\author[0000-0002-7743-2501]{Michael A. Keim}
\affil{Department of Astronomy, Yale University, New Haven, CT 06520, USA}

\author[0000-0001-9467-7298]{Allison Merritt}
\affil{Max-Planck-Institut fur Astronomie, Kunigstuhl 17, D-69117 Heidelberg, Germany}

\author[0000-0001-8367-6265]{Tim B. Miller}
\author[0000-0002-7075-9931]{Imad Pasha}
\author[0000-0002-5283-933X]{Ava Polzin}
\author[0000-0002-5120-1684]{Zili shen}
\affil{Department of Astronomy, Yale University, New Haven, CT 06520, USA}

\author[0000-0001-5310-4186 ]{Jielai Zhang \begin{CJK}{UTF8}{gbsn} (张洁莱)\end{CJK}}
\affil{Centre for Astrophysics and Supercomputing, Swinburne University of Technology, P.O. Box 218, H29, Hawthorn, VIC 3122, Australia}
\affil{Australian Research Council Centre of Excellence for Gravitational Wave Discovery (OzGrav), Australia}



\begin{abstract}
Uncertainty in the wide-angle Point Spread Function (PSF) at large angles (tens of arcseconds and beyond) is one of the dominant sources of error in a number of important quantities in observational astronomy. Examples include the stellar mass and shape of galactic halos and the maximum extent of starlight in the disks of nearby galaxies. However, modeling the wide-angle PSF has long been a challenge in astronomical imaging. In this paper, we present a self-consistent method to model the wide-angle PSF in images. Scattered light from multiple bright stars is fitted simultaneously with a background model to characterize the extended wing of the PSF using a Bayesian framework operating on a pixel-by-pixel level. The method is demonstrated using our software \texttt{elderflower} and is applied to data from the Dragonfly Telephoto Array to model its PSF out to 20--25 arcminutes.
We compare the wide-angle PSF of Dragonfly to that of a number of other telescopes, including the SDSS PSF and show that, on scales of arcminutes, the scattered light in the Dragonfly PSF is markedly lower than that of other wide-field imaging telescopes. The energy in the wings of the Dragonfly point-spread function is sufficiently low that optical cleanliness plays an important role in defining the PSF. This component of the PSF can be modeled accurately, highlighting the power of our self-contained approach. 
\end{abstract}



\section{Introduction}\label{Sec:Intro} 
Deep wide-field imaging has played an important role in studying diffuse, low surface brightness astronomical objects, such as ultra diffuse galaxies (e.g., \citealt{2015ApJ...798L..45V}), filaments or bubbles of the interstellar medium (e.g., \citealt{2000AJ....119.2919B}), tidal structures in the outskirts of galaxies (e.g., \citealt{2010AJ....140..962M}), gravitational lenses (e.g., \citealt{1996ApJ...466..623B}), Galactic cirrus (e.g., \citealt{1989ApJ...346..773G}), and intracluster light (e.g., \citealt{2005ApJ...631L..41M}). In many cases, ground-based and space-borne telescopes have been able to probe down to remarkably low surface brightness levels, but high-precision photometry of the revealed structures has proven challenging. One major source of systematic error is light in the outer wings of the stellar point-spread function (PSF), which we will refer to as the `wide-angle PSF'\footnote{Below we refer to the PSF beyond 30 arcsec as the `wide-angle PSF', although the practical definition depends on the scenario.}, although it is sometimes referred to as the `stellar aureole' \citep{1971PASP...83..199K}.

The central part (within a few tens of arcseconds) of the PSF is generally well understood and for ground-based telescopes it can be well represented by a Moffat function (\citealt{1969A&A.....3..455M}). \cite{1996PASP..108..699R} showed that this analytical form is consistent with predictions from Kolmogorov's atmospheric turbulence model (\citealt{1941DoSSR..30..301K}). However, on large scales, the PSF deviates from a single Moffat profile. The wide-angle PSF finds its origin in many processes, including propagation of the wavefront through the turbulent atmosphere, scattering from optical surfaces and optical inhomogeneities, and the properties of detectors. 
The pioneering work in this area was done by \cite{1971PASP...83..199K}, who was the first to collect and measure the extended wing of the PSF. Based on a joint profile spanning a range of 27 mag, King found that the wide-angle PSF is best described by an inverse-square law ($I \sim r^{-2}$, where $r$ is the radial angle on the sky). Subsequent measurements (e.g., \citealt{1973PASP...85..533K}, \citealt{2007ApJ...666..663B}, \citealt{2008MNRAS.388.1521D}, \citealt{2009PASP..121.1267S}, \citealt{2014A&A...567A..97S}, \citealt{2020MNRAS.491.5317I}) based on observations from different telescopes, in different bandpasses, and under different weather conditions mostly report a power-law-like form of the wide-angle PSF, but with a range of exponents:
$I_{p} \sim \,r^{-n}$,
with the power index $n$ ranging from 1.6 to 3 (\citealt{2014A&A...567A..97S}). It should be noted that these power-law profiles average over structures such as spikes or bumps caused by additional sources of diffraction and/or scattering (e.g., \citealt{2008MNRAS.388.1521D}, \citealt{2009PASP..121.1267S}). 

Even after over four decades of investigation, the physical origin of the wide-angle PSF is still unclear. As noted earlier, some studies propose that the wings of the PSF originate from the telescope and/or within the instrument (e.g., \citealt{1974ApJ...188..233S}, \citealt{1995seft.conf..303B}, \citealt{1996PASP..108..699R}, \citealt{2009PASP..121.1267S}). However, other studies argue that the main contribution arises externally, from scattering by thin atmospheric cirri, aerosols, or dust in the atmosphere (e.g., \citealt{2013JGRD..118.5679D}). These possibilities are not mutually exclusive and, as \cite{1996PASP..108..699R} noted, the wide-angle PSF may find its origin in a variety of sources.

In recent years, this topic of the wide-angle PSF has grown in importance, as it has become clear that it plays a central role in a number of important observations (see \citealt{2014A&A...567A..97S} for a review). Several studies have emphasized that mistreatment of the PSF wings could introduce systematic bias in the derivation of the \textit{intrinsic} light distribution of sources. One well-known example of one such bias is the `red halo' effect (\citealt{1998SPIE.3355..608S}, \citealt{2008MNRAS.388.1521D}) in Sloan Digital Sky Survey (SDSS) data. In this case, light at red wavelengths is scattered more than light at blue wavelengths, due to the use of a thinned CCD. This wavelength dependence of the PSF results in an artificial color gradient, which overshadows interpretations of physical color gradients (e.g., \citealt{2008MNRAS.388.1521D}, \citealt{2014A&A...567A..97S}, \citealt{2014MNRAS.443.1433D}). Systematic biases can also occur from using an overtruncated model PSF, which might cause underestimates of the total flux, false detection of disk flattening, and incorrect modeling of the ellipticity of stellar halos (e.g., \citealt{2008MNRAS.388.1521D}, \citealt{2014MNRAS.443.1433D}, \citealt{2014A&A...567A..97S}). This is especially important for analyses of edge-on galaxies, whose intrinsic brightness profiles can drop more sharply than the PSF itself (e.g., \citealt{2014A&A...567A..97S}, \citealt{2019MNRAS.483..664M}, \citealt{2020ApJ...897..108G}).

One important motivation for the construction of the Dragonfly Telephoto Array is to reduce scattered light in the PSF on large scales. Composed of well-coated and highly baffled telephoto lenses, Dragonfly is expected to have better-controlled PSF wings compared to reflecting telescopes, since reflective surfaces introduce more scattering within the telescopes (\citealt{2008SPIE.7012E..31N}). The PSF of Dragonfly was introduced in \cite{2014PASP..126...55A}, where a PSF on a degree scale was measured and found to have less energy in its outer wings than other wide-field survey telescopes. More recently, \cite{2020MNRAS.495.4570M} has shown that the PSF effects on measurements of galaxy outskirts in Dragonfly data are relatively small by convolving simulated galaxies with the Dragonfly PSF. However, this has not been tested in general cases because direct measurement of an `instantaneous' local wide-angle PSF is a non-trivial task.



In this paper, we present a self-consistent method for modeling the wide-angle PSF in deep images. Compared with classical methods for PSF measurement (e.g., doing azimuthally averaged photometry with $\chi^2$ fitting of individual bright stars), our approach models the full 2D image as a complete `scene'. The methodology is particularly well suited to crowded fields and ultra deep images where the stray light from stars contaminates {every pixel below a certain surface brightness floor (\citealt{2009PASP..121.1267S})}. Bayesian analysis allows us to incorporate prior knowledge about the extended wings of the PSF, and of the sky background, into the modeling. Most importantly, our method can be run on any image and hence provide the PSF appropriate for that specific observation.

In Section \ref{Sec:challenge}, we illustrate the biases inherent in classical profile measurement. In Section \ref{Sec:Bayesian}, we introduce the principles of Bayesian analysis and describe how our approach overcomes the biases described in Section \ref{Sec:challenge}. Section \ref{Sec:Data} briefly introduces the Dragonfly telescope, and in Section \ref{Sec:Methodology} we illustrate our methodology and present \texttt{elderflower}, a Python package for the implementation of our method. Section \ref{Sec:DF_PSF} presents a wide-angle PSF measured from Dragonfly data of M44, together with sanity checks. We also compare the PSF of Dragonfly with wide-angle PSFs of other telescopes, including the 2.5m telescope at the Apache Point Observatory used to undertake SDSS. Further illustrative applications of \texttt{elderflower} on Dragonfly data are presented in Section \ref{Sec:Case}, including investigation on the effect of dust on the lens. Section \ref{Sec:Discussion} discusses caveats, and Section \ref{Sec:summary} summarizes our results.

\section{Systematics of Profile Measurement at Low Surface Brightness Levels}  \label{Sec:challenge} 
    
This section illustrates the practical challenges when measuring the PSF at low surface brightness levels.\footnote{Similar considerations will apply when using azimuthal averaging to determine the radial profiles of faint sources, e.g., ultra diffuse galaxies.} 
    
\begin{figure*}[!htbp]
    \centering
      \resizebox{0.65\hsize}{!}{\includegraphics{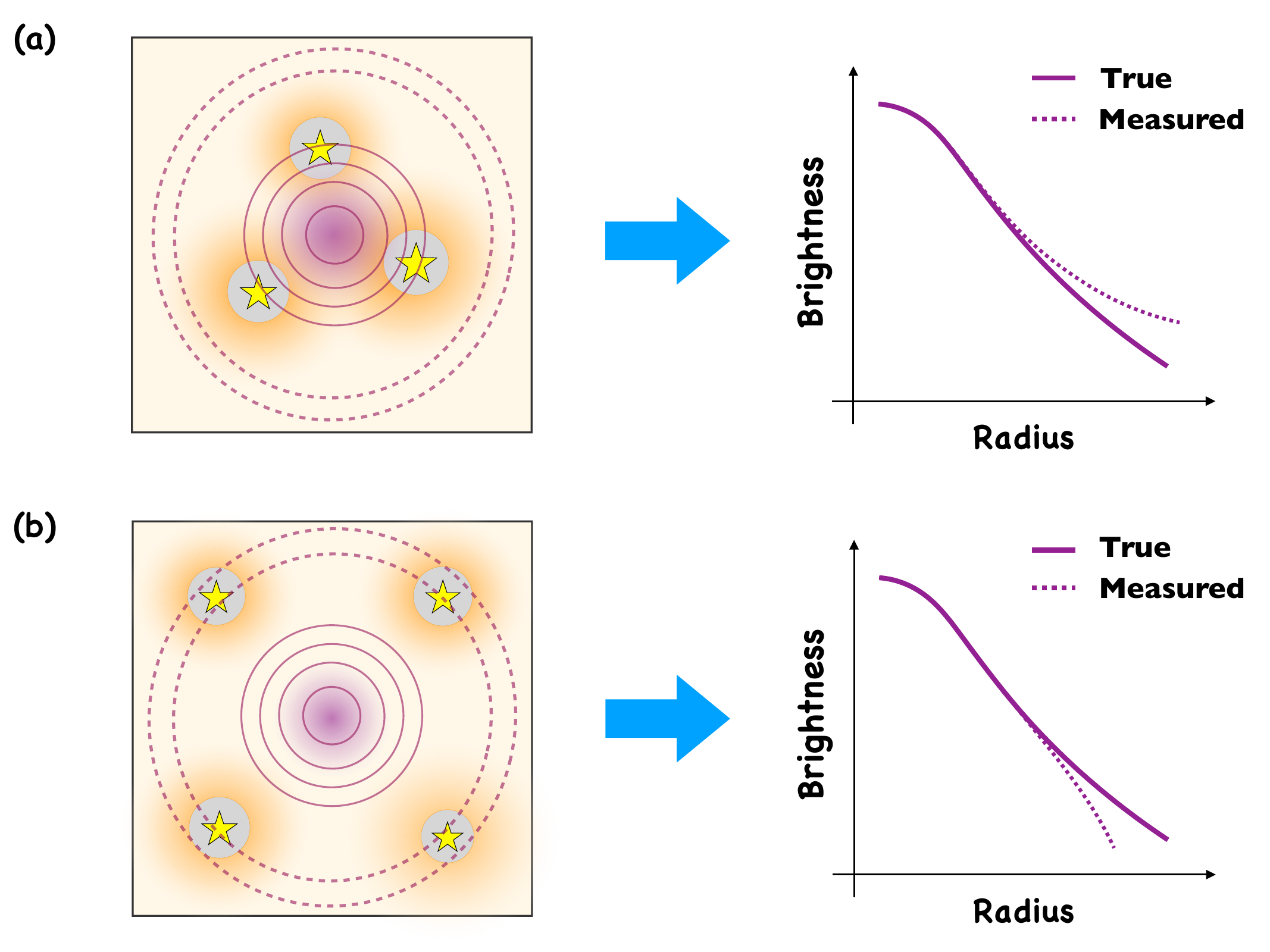}}
      \caption{A schematic diagram showing systematic errors in the measurement of the profile of a target object (purple blob) caused by bright stars and a biased sky at low surface brightness levels. The profiles are measured with the annuli in solid lines, and the sky values are measured by the ring in dashed lines. The two cases shown illustrate how artificial flattening and/or steepening can both emerge, depending on the geometry and brightness of the contaminating sources. (a) In this case, the geometry of contaminating starlight allows the sky level to be accurately estimated, so that `leaking' light from bright stars is assigned to the target, which causes a flattening in the measured profile.   (b) In this case, azimuthal averaging results in an over-estimated sky value and thus an artificial steepening in the object's profile. In both cases, heavy masking could be used in an attempt to mitigate these issue but at the cost of reducing the number of pixels for analysis. As described in the text, this crude approach has significant limitations.}
    \label{fig:schematic}
\end{figure*}

The traditional approach for measuring the wide-angle PSF is composed of the following steps: (1) selection of a few bright (ideally isolated) stars; (2) definition of a series of annuli for each star (or alternatively, using box estimators, e.g., the mode estimator in SExtractor) to determine the local backgrounds; (3) subtraction of the background from the data; (4) measurement of azimuthally averaged light profiles in radial bins with nearby sources masked; and (5) averaging (or interpolation) over individual stars to map out spatial variations in the PSF. Caution needs to be exercised when applying these steps to deep images down to low surface brightness levels, because some key assumptions, such as the possibility of defining the sky level independently for all sources, might break down. A schematic illustration showing some of the problems inherent in this approach is presented in Figure \ref{fig:schematic}. This figure illustrates the basic point that, at low surface brightness levels, the shape of the profile becomes uncertain and can easily become modified by the threshold used for masking, the evaluation of the background, the spatial distributions of bright sources, and other factors.

If  scattered light is indistinguishable from the local background, sky background estimates will be biased high. In extreme cases, scattered light permeates the entire field of view, such as in the image of the central region of the Virgo Cluster by \cite{2009PASP..121.1267S}, a pioneering work that uses ray tracing to model the scattered light. Understanding the role of biased sky measurements is crucial to obtaining reliable galaxy profiles in such fields, as has been pointed out (and partially tackled) in many studies (e.g., \citealt{1987PASP...99..191S}, \citealt{2017MNRAS.467..490F}, \citealt{2018MNRAS.475.3348H}, \citealt{2018PASP..130h4504J}). For example, \cite{2017MNRAS.467..490F} re-estimated the sky by fitting it simultaneously with objects for SDSS data and found the sky values around objects with large sizes or luminosities are overestimated by the SDSS pipelines. When doing photometry, adopting an over-estimated sky, e.g., caused by scattered light of the target itself, imposes unphysical steepening/truncation on the measured profile and light loss in the derivation of the total luminosity. Of course, depending on the purpose of the investigation, accuracy in the profile at low surface brightness levels might or might not be important.
    
Attempts to deal with sources of scattered light often center upon masking to eliminate sources of contamination. This approach has important limitations (e.g. \citealt{1987PASP...99..191S}, \citealt{2018PASJ...70S...7C}, \citealt{2018MNRAS.481.3170M},  \citealt{2019A&A...625A..36W}). The main problem is leakage of light from the extended wings of the PSF. The issue is particularly significant at ultra low surface brightness levels and/or in crowded fields. In these cases, the outskirts of measured profiles suffer from artificial flattening.
The typical level of masking, e.g., based on a signal-to-noise ratio (S/N) threshold of 3, is insufficient. Very aggressive masking can \textit{mitigate} the situation but does not \textit{solve} it. As the surface brightness limit goes down, the field becomes so heavily masked that one effectively runs out of pixels for analysis. 

The discussion thus far has focused on systematics that complicate sky estimates, but another factor that must be addressed when measuring the wide-angle PSF is the need to obtain high S/N profiles at large angles, which can be difficult to achieve in fields that are devoid of bright stars. An approach that has been used in large surveys (e.g., SDSS: \citealt{2008MNRAS.388.1521D}, \citealt{2011ApJ...731...89T}, \citealt{2020MNRAS.491.5317I}, DES: \citealt{2019ApJ...874..165Z}) is to register, scale, and stack a large number of images of (ideally) isolated stars. This stacking technique may not account for large-scale spatial variations across the field, leading to a deviation from the truth. If an insufficient number of stars is available on a target image, multiple epochs would need to be used to obtain the required signal in the wings of the PSF, which would leave out their temporal variations.

The complications noted above have led authors to explore a variety of different approaches to measuring the PSF (e.g., \citealt{2007PASP..119.1403J}, \citealt{2007A&A...461..373M}, \citealt{2012MNRAS.419.2356B}, \citealt{2013ascl.soft01001B}, \citealt{2017A&A...601A..86K}, \citealt{2018JCAP...07..054H}, \citealt{2020MNRAS.496.4209F}, \citealt{2021MNRAS.501.1282J}). Some of these have led to great improvements in estimates of the PSF on small angular scales, but thus far few have attempted to deal with the problem of the wide-angle PSF on scales of tens of arcmin. The lack of attention to the wide-angle PSF is acceptable when it is not the dominant systematic for the science being investigated. However, for low surface brightness science, e.g., outskirts of nearby galaxies, the wide-angle PSF can indeed be the dominant systematic, and new approaches are needed (\citealt{2014A&A...567A..97S}).  

In the present paper, our proposed solution is to forward model the background and \textit{all} sources simultaneously. This accounts for the fact that  scattered light from multiple sources is coupled, and this scattered light cannot be treated as independent from the `true' sky.
Because all the scattered light is fully accounted for, no exhaustive masking is required, although for practical reasons some compromises (such as mild masking) are required. We present our ideas and an implementation of our methods in the following sections.

\section{Bayesian PSF Modeling}  \label{Sec:Bayesian} 
    
Before going into the details of our methods, we introduce the Bayesian framework of PSF modeling and describe how it mitigates relevant systematics. Readers familiar with Bayesian modeling may skip this section.

In the past two decades, Bayesian analysis has become a very popular tool for data analysis in astronomy. Given a reasonable likelihood model, Bayesian analysis incorporates a priori knowledge on parameters into the model to yield a posterior distribution. The fundamental Bayes theorem states that the posterior is given by: 
    
\begin{equation}
    p({\bf\Theta}|D,M)=\frac{p(D|{\bf\Theta},M) \cdot p({\bf\Theta}|M)}{p(D|M)}\,,
\end{equation}
where $p(D|{\bf\Theta},M)$ is the likelihood of the data ($D$) given the parameters ($\bf\Theta$), $p({\bf\Theta}|M)$ is the prior (known information), and ${p(D|M)}$ is the evidence (marginal likelihood). Evidence can be used to compare different models by comparing the ratio of their evidence. 

PSF modeling under the Bayesian framework can therefore be described as a process of: (1) building priors and models of the PSF and the sky; (2) generating images based on models with proposed parameter sets; (3) calculating their likelihoods by comparing the model and the data; and (4) combining the above to obtain posterior probabilities of parameters.
    
A central aspect in modern Bayesian analysis is posterior sampling, of which Monte Carlo Markov Chain (MCMC) is the most commonly used approach, with much progress having been made in this area (\citealt{2017ARA&A..55..213S}). Alternatively, nested sampling has been proposed (\citealt{2004AIPC..735..395S}) to overcome some of the issues confronting MCMC. Nested sampling generates samples from nested shells of prior volume with increasing likelihood. It then estimates the posterior using weights of samples. The major advantages of nested sampling over MCMC include, for example, that it directly returns the evidence for model comparisons, it has a definitive stopping criterion, and it is able to increase the sampling efficiency in specific types of problems. We refer to  \cite{2004AIPC..735..395S} and \cite{2020MNRAS.493.3132S} for its detailed algorithm, advantages, and limitations. This work adopts nested sampling as the sampling method.
    
Our approach simultaneously models multiple sources (whose positions and magnitudes are known in advance), convolved with a PSF defined on scales out to tens of arcminutes, along with a background that has a large-scale spatial variation. This approach explicitly incorporates scattered light from large areas of the image, which is the major origin of biased background estimates, and accounts for the coupling of scattered light and background estimates. In some aspects, our approach is similar to that applied in \cite{2014MNRAS.443.1433D}, who used related ideas to model galactic halos instead of attempting to model the PSF. 

\section{The Dragonfly Telephoto Array}  \label{Sec:Data}
 
To make these ideas concrete, we will illustrate our approach using deep wide-field imaging data from the Dragonfly Telephoto Array (\citealt{2014PASP..126...55A}). 

Dragonfly is an array composed of 48 Canon 400 mm $f/2.8$ IS II USM-L telephoto lenses. Its design is optimized for detecting faint, diffuse light on large scales (from a few arcmin to several degrees) down to an ultra-low surface brightness ($1\sigma$ at $\sim$30.5 mag/arcsec$^2$). Half of the cameras image in the Sloan $g$-band and the other half in the Sloan $r$-band. The fields of view of individual lenses are slightly offset such that the total field of view of the array is about $2\degree \times 3\degree$ with a pixel scale of 2.85\arcsec/pix. The final images produced by the array are stacks of hundreds or thousands of short exposure (10 min) frames resampled to a pixel scale of 2.5\arcsec/pix. Detailed descriptions of the telescope and the data reduction pipeline can be found in \cite{2014PASP..126...55A} and \cite{2020ApJ...894..119D}.
 
Several aspects in the design of Dragonfly make it less affected by systematics caused by scattered light on large scales: (1) no use of reflective optical surfaces, which reduces backscattering into the optical path by reflection; (2) the use of nano-fabricated coatings, which suppresses internal reflection; (3) no obstruction in the pupil, which reduces energy spread at large scales from diffraction. In 2014, a white light image of Venus was used to measure the Dragonfly PSF to prove the effectiveness of these concepts (\citealt{2014PASP..126...55A}). The measured PSF is well-behaved out to large radius with a suppressed extended wing and does not show strong diffraction spikes, strong ghosts, or other high-order features. 


\section{Methodology}  \label{Sec:Methodology} 

\begin{figure*}[!htbp]
\centering
  \resizebox{0.85\hsize}{!}{\includegraphics{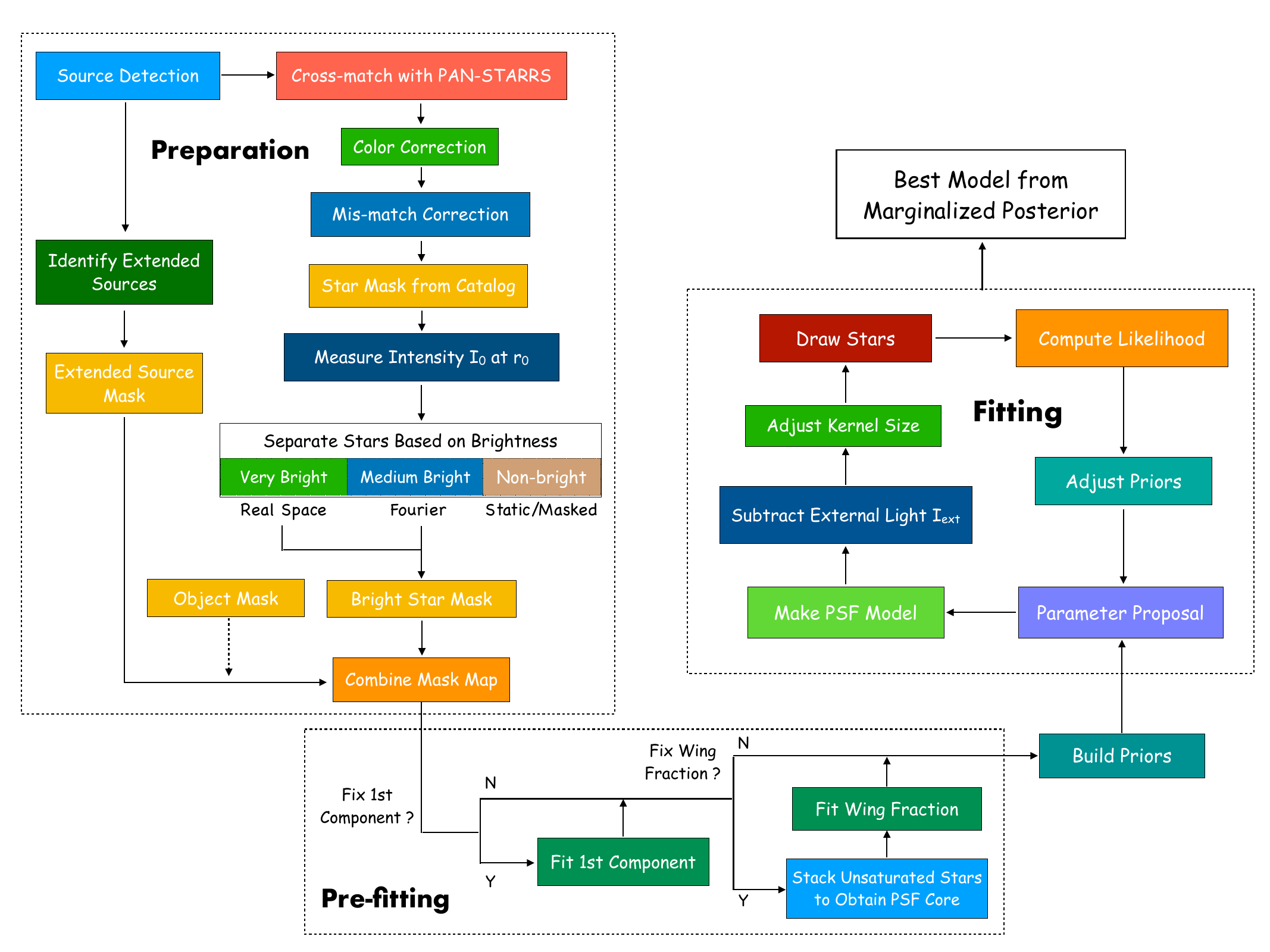}}
  \caption{Workflow of our PSF modeling implemented in \texttt{elderflower}. The procedures consist of three sections: the preparations for fitting, an optional maximum likelihood fitting for partial constraints, and the full Bayesian fitting with the forward model. The final PSF is constructed with a core from stacking unsaturated stars and a parametric wing from the fitting.}
\label{fig:workflow}
\end{figure*}

This section describes our methodology for the modeling of the wide-angle PSF in deep wide-field images. In brief, we assume the PSF is a combination of a Moffat core component and a multi-power law wing component, and simultaneously fit all bright objects in the field with the PSF under the Bayesian framework. Below we specify the algorithm of the method. The procedures are summarized in Figure \ref{fig:workflow} and implemented in the Python package, \texttt{elderflower}. {The software \texttt{elderflower} is available on GitHub\footnote{ \url{https://github.com/NGC4676/elderflower}}, and the version used at the time of this work is archived in Zenodo \citep{elderflower}.} Although the software tool is developed for the Dragonfly Telephoto Array, it is compatible with other wide-field observations and the core idea is transferable.

\subsection{Source Photometry} \label{Sec:prep}

For the construction of point source models, the primary elements are positions and scalings/normalizations. These are obtained from stellar photometry done internally or from public catalogs. Extended sources will be masked in the image. We only model stars brighter than a certain limit, since extended wings from these are dominant. Including faint stars whose extended wings are far below the background noise would have negligible information benefit but dramatically increase the running time. On the other hand, due to the large pixel size of Dragonfly, faint stars are not well sampled; so completely ignoring them leaves extra light/bias on the background modeling. In practice, we group stars by brightness and use different treatment in terms of normalization and masking. We will describe the approach used for Dragonfly data, but the best strategy for generating the point source models could be different for other telescopes. 
    
    \subsubsection{General Star Catalog Considerations} \label{Sec:general_consideration}
    
    We start with an initial run of source detection using SExtractor (\citealt{1996A&AS..117..393B}). The detection and analysis threshold is set to be high with an S/N of 5 to pick out well-sampled bright sources. A segmentation map and a SExtractor catalog are generated. For each source, SExtractor computes a photometric magnitude (\texttt{MAG\_AUTO}) and positions that are estimated from image moments. However for the bright end of saturated stars (\texttt{MAG\_AUTO} $\lesssim$ 12 mag in typical Dragonfly data) the luminosities are underestimated and the centroids are also affected. Therefore, an external star catalog is used to supply information on bright stars: the SExtractor catalog is crossmatched with the Panoramic Survey Telescope and Rapid Response System (Pan-STARRS) catalog (\citealt{2016arXiv161205560C}) to utilize the cataloged positions and magnitudes. Thanks to the flexibility and exquisite optimization of its photometric measurement pipeline (\citealt{2020ApJS..251....6M}), Pan-STARRS has superb spatial location precision and dynamic range. 
    
    Bright stars are characterized in the Pan-STARRS \texttt{MeanObject} catalog \footnote{In the Pan-STARRS (\href{https://panstarrs.stsci.edu}{https://panstarrs.stsci.edu}) \texttt{MeanObject} catalog the photometric information of bright stars are measured by fitting flexible PSF models on radial profiles of single epoch data with logarithmically interpolated cores. {We use the mean PSF magnitudes from the catalog}. One drawback of using the \texttt{MeanObject} catalog is that very faint stars around the single epoch detection limit are missed or biased compared with the \texttt{StackObject} catalog. However, as stated before, these stars are not our main concern and are not well sampled in our data. We constrain the faint limit of crossmatching to be 22 mag in the \texttt{MeanObject} catalog. No bright limit is applied.}. 
    While both Dragonfly and Pan-STARRS images are obtained using SDSS filters, small color terms (from minor transmission curve differences and/or detector quantum efficiency variations across the filter bandpasses) must still be accounted for. Therefore, A color correction is computed between the Dragonfly band and the corresponding Pan-STARRS band using non-saturated stars. Below we refer to the magnitude corrected from the cataloged magnitude as $m_{corr}$. Thanks to the dynamical range of Pan-STARRS, $m_{corr}$ provides accurate photometry for stars $\gtrsim$ 10 mag.
    
    The procedure described works well for the vast majority of stars, but several concerns remain when using the cataloged photometric information of some luminous saturated stars ($m_{corr}$ $\lesssim$ 8-9 mag). Firstly, there are occasional multiple detections in the vicinities of their centers in the catalog which involves bad fitting in a single epoch, fainter stars around the center, spurious instrumental features, etc. Secondly, the measurements of their magnitudes (and possibly positions), might not be sufficiently accurate for a variety of reasons (e.g., saturation, bias in the background estimate). Using the total flux or magnitude $m_{corr}$ as normalization would potentially lead to incorrect scaling. Indeed, fixing the normalization breaks the coupling between the scattered light from stars and the sky. The normalization recipes are described in the following sections.
    
    In addition, among the saturated stars, a few are not properly measured and recorded in the Pan-STARRS catalog. Around 5\% of saturated stars in the SExtractor detections (excluding extended sources) are {mismatched} in the crossmatching with Pan-STARRS. This could be due to faint sources in the vicinity of the center of the star or inaccurate measurements in Pan-STARRS. 
    To properly include these stars in the sample, we fit an empirical relation between \texttt{MAG\_AUTO} and $m_{corr}$ using a piecewise linear function, {where the saturation occurs at the break}. If a source has $m_{corr}$ significantly deviated (with a difference of over 2 mag) from the expected value according to the empirical relation, it is added to the sample using the empirically corrected of \texttt{MAG\_AUTO}.

    \subsubsection{Very Bright Saturated Stars} \label{Sec:VB_star}
    
    The wide-angle PSF is largely derived from bright, saturated stars in the field.  A large fraction of the total light of these stars is in the nonlinear regime of the detector. In a typical 10 min exposure Dragonfly frame, saturation occurs for stars brighter than around 13.5 mag. To account for them, we first define a sample of `very bright' stars (denoted as `VB' stars) based on a user-defined threshold. The threshold is typically set between 10 and 12 mag, depending on the crowdedness of bright stars in the region. The number of VB stars in a region to be simultaneously modeled is controlled for computational efficiency.
    
    Getting proper normalization for bright, saturated stars is a non-trivial task. The flux measured from isophote/aperture photometry is usually biased. Saturation is one of the primary obstacles, besides high-order artifacts such as spikes, and tracking errors. Another concern is that the extended PSF wing also contributes to the total light. The fraction of total flux in the wing is uncertain without assumption on the PSF wing. Given the possible variation in the extended wing, adopting a fixed flux correction \textit{before} the modeling is not ideal. Furthermore, as stated in Section \ref{Sec:challenge}, the background value needs to be determined in advance, which, however, can be biased. Even if the background is well determined, the aperture flux still contains external light from other nearby sources. The effect would be prominent in the presence of a multitude of luminous stars in a small region of sky.
    
    For this purpose, we consider an approach based on the intensity at a certain radius. The intensity is measured and used to compute a normalization for each star according to the proposed PSF and background level \textit{during} the modeling, where external light from other stars is mutually subtracted through iteration. The validity of such normalization relies on the credibility of the model (see Section \ref{Sec:Discussion}). Although one could possibly measure and correct the flux during the modeling, it is much more computationally expensive to do photometry {of many stars} in each model proposal, and it is challenging to account for the external light.
        
    To normalize the very bright (VB) stars, we measure the $3\sigma$-clipped azimuthally averaged intensity $I_0$ at a scale radius, $r_0$. For each star, the external light from bright sources and a proposed local sky are subtracted:
    \begin{equation}
        \hat{I}_{*,i} = I_{0,i}-\Sigma_{j\,(\neq i)}\,I_{j}\,(\vec{r_{ij}}) - I_{sky}\,,
        \label{Eq:I_int}
    \end{equation}
    where $\hat{I}_{*,i}$ is an estimate of the scaling for the star $i$ in the proposal, and $\vec{r_{ij}}$ is the separation vector from star $j$ to star $i$. The normalization is computed over a few iterations. The fractional change is typically small ($<$0.1\%) after three to five iterations. For computational efficiency, the external light is only evaluated at the center of the star. This assumption only works on large scales, which may not be valid for close or blended sources, such as binary stars. {The radius $r_0$ is chosen to keep away from the saturated core of the brightest star in the region while keeping $I_0$ at least $5\sigma$ above an (estimated) background.} For Dragonfly, $I_0$ is measured using a thin annulus at $\sim$30\arcsec. When measuring $I_0$, nearby sources are masked using the mask map generated in Section \ref{Sec:general_consideration} as well as image artifacts. $\hat{I}_{*,i}$ then serves as the scaling factor for star $i$.
    
    To mask the saturated and undersampled stellar cores, a user-defined aperture radius is set. The aperture needs to be larger than the saturation area of the brightest star in the field of view. In addition, stellar spikes are masked for VB stars\footnote{Note that the spikes seen on Dragonfly data do not originate  in diffraction from a secondary mirror support structure; instead, they are very low-level artifacts from the micro lens array on the sensor.}.
    
    \subsubsection{Medium Bright Stars}
    For computational efficiency, a second sample of `medium bright' stars  (denoted as `MB' stars), is defined between the VB star magnitude limit and the magnitude at which saturation occurs ($\sim$13.5 mag for Dragonfly). This sample is rendered on the image using a different approach from VB stars (see Section \ref{Sec:realization}). 
    
    The normalization treatment for MB stars is similar to that for VB stars, except that scattered light from MB stars is not mutually subtracted to avoid $O(n^2)$ computation. {The scattered light from VB stars is still accounted for in the iteration as described in the previous section.} This computational shortcut is justified because light from VB stars takes up the majority of the total scattered light. The process of measuring the normalization of MB stars is the same as in Section \ref{Sec:VB_star}. The cores of MB stars are also masked by user-defined apertures. Because diffraction artifacts are not prominent for MB stars, there are no additional masks for them.

    \subsubsection{Intermediate Brightness and Faint Stars}
    
    Intermediate brightness stars, which are brighter than the faint limit (set as 15 mag for Dragonfly data) but fainter than the threshold for the MB stars, are rendered as a fixed component in the image during the fitting with a single power law. The parameters are fixed with fiducial values from stacking bright unsaturated stars. 
    
    Faint stars are not included in the modeling because their extended wings quickly merge into the sky with minimal influence on scales larger than 30\arcsec. Stars fainter than the faint limit are not modeled and are masked as follows. 

    Dragonfly's poor resolution means a large number of faint stars are unresolved and will be missed in the source detection process, and accordingly in the crossmatching to the Pan-STARRS catalog. A mask map is built from the full Pan-STARRS catalog down to 22 mag conservatively for the general purpose of masking stars. Stars fainter than 22 mag are considered as part of the sky. An empirical relation between the object Kron radius and magnitude is obtained with the SExtractor segmentation map by a $3\sigma$-clipped second-order polynomial fitting using sources of $13.5<m_{corr}<22$. The empirical size is then enlarged for a slightly more aggressive mask than the input. Typically, the mask range of a source at 15 mag is out to around 30\arcsec. The faint end has a minimum aperture size of 5{\arcsec}. Masks of MB and VB stars are replaced with aperture masks as described in previous sections.

    \subsubsection{Bright Extended Sources}
    
    Extended sources are currently not modeled and are masked empirically. We pick out extended sources based on a relationship between peak surface brightness and magnitude. A broken power law is iteratively fitted with $3\sigma$ clipping for the relationship between the central maximum brightness \texttt{MU\_MAX} and the magnitude \texttt{MAG\_AUTO} from SExtractor. Sources with $5\sigma$ deviation from the relation and either  \texttt{CLASS\_STAR} $<$ 0.5 or \texttt{ELLIPTICITY} $>$ 0.7 are deemed to be extend sources and are masked with elliptical apertures built from SExtractor shape parameters but with enlarged aperture sizes. This step is designed to maximally mask the light from sources not included in the model, although we note there could be other sources of contamination (see Section \ref{Sec:Discussion}).
    In addition, a user-defined mask map can be optionally provided to mask particular regions/objects of interest (referred to as `object mask'), e.g., diffuse galaxies under the detection threshold, large nearby galaxies, subregions filled with Galactic cirrus, etc., as supplementary masks. The final mask map is the combination of the above introduced mask maps.

\subsection{Likelihood Function} \label{sec:likelihood} 
    We denote the parameter set as $\bf \Theta$, which characterizes the shape of the PSF ($I_{\rm PSF}$) and the sky background ($I_{\rm sky}$), given the position $\{x_{*,i},y_{*,i}\}$ and the normalization $\hat{I}_{*,i}$ of each star\footnote{In principle one could move $\{x_{*,i}, y_{*,i}\}$ into $\bf \Theta$, i.e. treating the star positions as parameters. However, it would dramatically increase the computational complexity as the number of modeled stars increases. Small offsets in the positions of stars are likely to have negligible effects on large scales.}. The variables are pixel-based, i.e. dependent on positions (x, y) of an unmasked pixel. The intensity (or surface brightness) of a pixel (x, y) is given by:
    \begin{align}
    I\,(x,y | {\bf \Theta}) \,= \,&\Sigma_i \,F_i \ast I_{\rm PSF}\,(x, y | {\bf \Theta}, x_{*,i},y_{*,i}, \hat{I}_{*,i}) \nonumber\\ 
            &+ I_{\rm sky}(x,y | {\bf \Theta}) \,,
    \label{Eq:I_tot}
    \end{align}
    where $F_i$ is a $\delta$-function at the position of each star multiplied by its total flux, $\ast$ stands for convolution, and $\Sigma_i$ denotes the sum for bright stars in the model. 
    
    The PSF is assumed to vary with radius and is modeled as the sum of the core component described by a Moffat function $I_{\rm M}(\theta)$ and the aureole component $I_{\text {p}}(\theta)$ described by a multi-power law:
    \begin{align}
    I_{\rm PSF}(\theta) =\, & (1-f_{\rm p}) \,\cdot\, I_{\rm M} (\theta|\gamma, \alpha)  \nonumber\\ 
            &+ f_{\rm p} \cdot I_{\rm p}(\theta|n_0, n_1,...,n_k,\theta_1,..., \theta_k)\,,
    \end{align}
    where $f_{\rm p}$ is the flux fraction of the multi-power law component, and $n_k$ and $\theta_k$ are the power law index and transition radius of the k-th component. The power law is flattened within the inner $\theta_0=$ 5{\arcsec} for numerical stability. 
    
    The shape of the central Moffat function {(normalized to 1)} is defined by $\gamma$ and $\alpha$:
    \begin{align}
    I_{\rm M} (\theta) = \frac{\alpha-1}{\pi\gamma^2} \left(1 + \frac{\theta^{2}}{\gamma^{2}}\right)^{- \alpha}
    \end{align}
    The core component of the PSF drops sharply with radius, and falls within the mask. Since they are independent from the aureole component in the model, the parameters of the inner PSF ($\gamma$, $\alpha$) are kept fixed during the fitting. The values are fitted from the core parts of the stacked profile from bright non-saturated stars. 
    
    The multi-power law component is defined by:
    \begin{align}
    I_{\text {p}}(\theta)= \sum_{l=0}^{k} \delta_{lk} A_{l} \cdot \theta ^{-n_{l}}\,,
    \end{align}
    {where $\delta_{lk}=1$ if $l=k$ and otherwise $\delta_{lk}=0$.} The normalization of each sub-component is given by the scaling factor $\hat{I}_{*,i}$ measured at scale radius $r_0$:
    \begin{align}
    \mathrm{A}_{l}=\left\{
    \begin{array}{ll}
    ({\hat{I}_{*,i}\ r_{0}^{n_{0}}}/{f_p F_i})\cdot {\displaystyle  \prod_{s=1}^{l}}\left(\theta_{s}^{n_{s}} / \theta_{s}^{n_{s-1}}\right) & (l \geq 1) \\
    ({\hat{I}_{*,i}\ r_{0}^{n_{0}}}/{f_p F_i}) & (l = 0)
    \end{array}\right.
    \label{Eq:norm_pow}
    \end{align}
    {Note $F_i$ and $f_p$ in Eq. \ref{Eq:I_tot}--\ref{Eq:norm_pow} cancel out and thus do not affect the normalization of $I_{\text {p}}(\theta)$.} In the wide-angle range that we are concerned with, the power law component dominates over the Moffat component.
    
    {The smoothly varying sky background $\mu_{\rm sky}$ is represented by a low-order bivariate Legendre polynomial to account for background structure not introduced by scattered light from wide-angle PSF of stars:}
    \begin{equation}
    \mu_{\rm sky} = \mu + \Sigma_{ij}\,c_{ij}\,P_{ij>0}(x,y)\,,
    \end{equation}
    where $\sigma_{\rm sky}$ is the standard deviation of sky\footnote{{In principle, $\sigma_{\rm sky}$ can be calculated from the pixel counts based on Poisson noise. In the case of a background level having been removed in the reduction pipeline (as is the case for the Dragonfly images used below), we estimate $\sigma_{\rm sky}$ by treating it as a free parameter in the analysis.}},
    and $c_{ij}$ is the coefficient of the $ij$-th 2D Legendre function $P_{ij}$. 
    Because we are modeling a relatively small area of sky compared to the full field of view, we choose to conservatively limit Legendre polynomials to first order.
    The sky background is then modeled as random variables drawn from a Gaussian distribution:
    \begin{equation}
    I_{\rm sky}(x,y) \sim \, \mathcal{N}(\mu_{\rm sky},\, \sigma_{\rm sky})
    \end{equation}
    
    The scaling of each star $\hat{I}_{*,i}$ is then converted into the total flux $F_i$ by integrating $I_{\rm PSF}$ to infinity. Since $I_{\rm PSF}$ is assumed to be independent of individual stars, $\hat{I}_{*,i}$ can be moved outside of the integral, which makes the conversion become a factor that can be analytically derived:
    \begin{equation}
    \begin{split}
    F_i = \int_0^\infty I_{\rm PSF}(\theta| {\bf \Theta}) \hat{I}_{*,i}({\mu_{\rm sky}}) \, d\theta = C({\bf \Theta}) \hat{I}_{*,i}({\mu_{\rm sky}})
    \end{split}
    \end{equation}
    
    The log-likelihood can be written as:
    \begin{equation}
    \begin{split}
    \ln{L({\bf \Theta})} = & - \frac{1}{2} \left[I({\bf \Theta})-\hat{I} \right]^T\Sigma^{-1}\left[I({\bf \Theta})-\hat{I} \right] \\
                           & - \frac{1}{2} \log \left[(2\pi)^N |\Sigma| \right]\,,
    \end{split}
    \end{equation}
    where $N$ is the number of unmasked pixels, $\Sigma$ is the $N\times N$ covariance matrix, and $\hat{I}$ is the data image after masking. 
    
    With the assumption of independently and identically distributed noise, $\Sigma$ only has diagonal elements as the pixel noise:
    
    \begin{equation} \label{eq:uncertainty}
    \sigma_k = \sqrt{\sigma_{\rm sky}^2 + (\hat{I_k}-I_{k,{\rm sky}})/g}\,,\,k=1,2,...,N\,,
    \end{equation}
    where $\hat{I_k}$ is the intensity of the $k$-th pixel, $I_{k,{\rm sky}}$ is the proposed sky background at the pixel position, and $g$ is the effective detector gain.

\subsection{Priors} \label{sec:prior}
    The next step is to construct priors of parameters. The full set of free parameters for the fitting is: ${\bf \Theta} = \{n_0, n_1, ..., n_k, \theta_1, ..., \theta_k, \mu, \sigma_{sky}, c_{10}, c_{01}, f_{p} \}$, which is summarized in Table \ref{table_prior}. Depending on different applications, a subset of the parameters can be well constrained separately from others.
    
    \begin{table*}
    \centering
    \caption{Parameters and the Corresponding Priors Used in the Modeling.} \label{table_prior}
    \begin{tabular}{ccc}
    \hline\hline
    Parameter & Description & Prior \\
    \hline
    $n_0{\ }^{\dagger}$ & Power law index of the first power law component & $\mathcal{N}(\hat{n_0}, \delta n_0)$ \\
    $n_{k+1}$ & Power law index of the [k+1]-th (k$\geq$0) power law component & ${U}[{\rm max}\{n_{\rm min},\, n_k-\Delta n\}, n_k-\delta n]$ \\
    $\log\theta_1$ & Transition radius (in arcsec) of the first power law component  & ${U}[\log\theta_{\rm min},\, \log\theta_{\rm max}]$ \\ 
    $\log\theta_{k+1}$ & Transition radius (in arcsec) of the [k+1]-th (k$\geq$0) power law component & ${U}[\log\theta_{k}, \log\theta_{\rm max}]$ \\
    $\mu$ & Mean of the sky background & $\mathcal{N}(\hat{\mu}, \hat{\sigma}),\, \mu \leq \hat{\mu}$ \\
    $\sigma_{\rm sky}{\ }^{\dagger}$ & Standard deviation of the sky background & $\log\mathcal{N}(\hat{\sigma}, 0.3),\, \sigma_{\rm sky} \leq \hat{\sigma}$ \\
    $c_{ij}$ & Coefficients of the [i+j]-th order Legendre Polynomials in the sky background & ${U}[0, \hat{\sigma}]$ \\
    $\log\,f_{p}{\ }^{\dagger}$ & Fraction of light in the power law component(s) & ${U}[0.01, 0.4]$ \\
    \hline
    \end{tabular}
    \tablenotetext{\dagger}{Parameters that are optionally constrained by maximum likelihood fitting before the Bayesian fitting.}
    \end{table*}
    
    The PSF of Dragonfly is smooth over a wide dynamic range, and its outer portion can be well described by power laws. Several power law components are defined as described in Table \ref{table_prior}. 
    
\begin{figure}[!htbp]
    \centering
      \resizebox{\hsize}{!}{\includegraphics{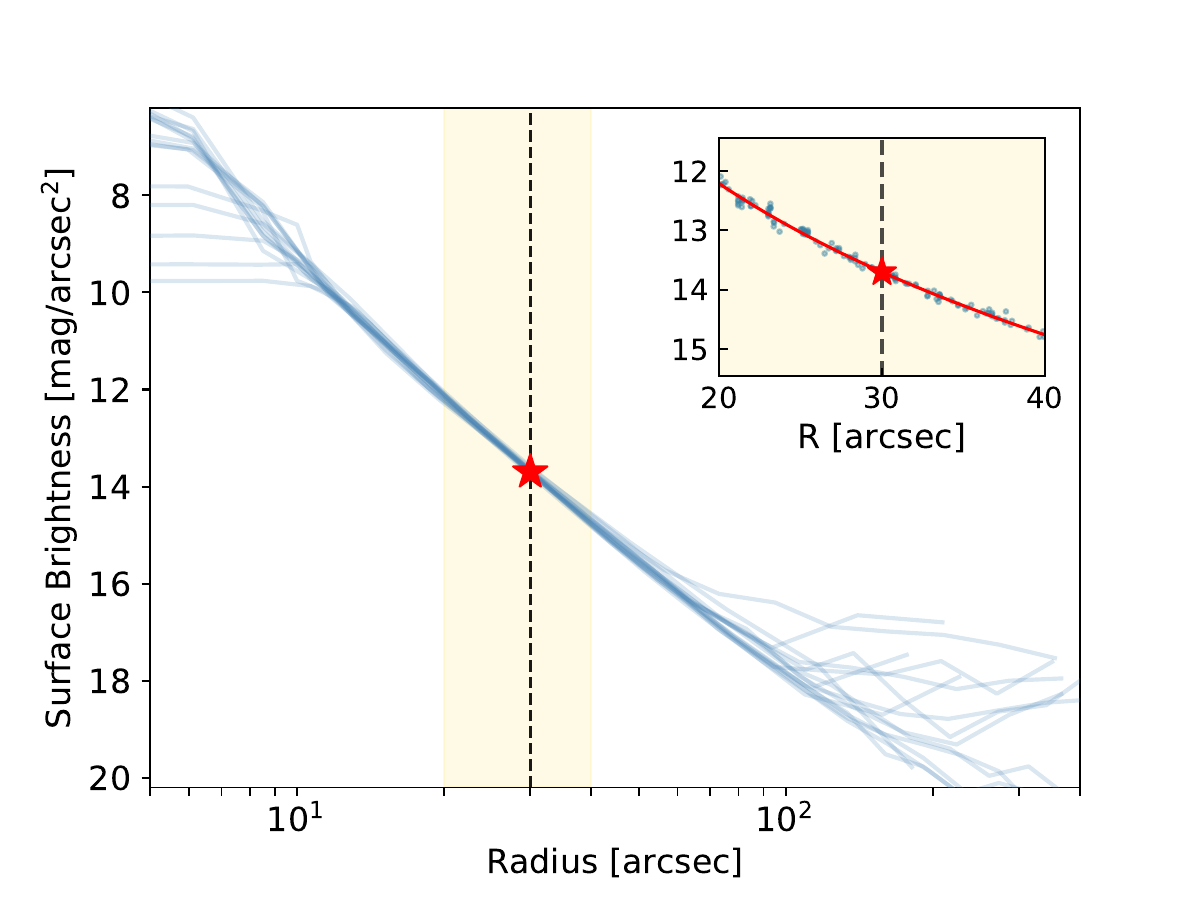}}
      \caption{Illustration of fitting the first component showing 1D azimuthally averaged radial profiles of the 15 brightest stars normalized at $r_0$ = 30{\arcsec} from a cutout in the NGC 5907 field (see Section \ref{Sec:ngc5907}). The first power law component is fitted using individual stellar profiles within the range $20{\arcsec}-40{\arcsec}$. The normalization is indicated by the red star and its position by the black dashed line. The profiles are normalized to have the same surface brightness that corresponds to a magnitude 0 star based on the final model PSF. The insert panel shows a zoom-in with the x-axis on a linear scale.}
    \label{fig:n0_ngc5907}
    \end{figure}

    The first power law component can be well estimated/constrained from stacking 1D profiles of several bright stars above 5$\sigma$ in the field. To achieve this, we run \texttt{elderflower} in a mode where the degeneracy {between $\hat{n_0}$ and other parameters} is slightly broken by performing a maximum likelihood fitting before running the Bayesian fitting. An estimated value $\hat{n_0}$ and its uncertainty $\delta \hat{n_0}$ are obtained using sigma-clipped azimuthally averaged profiles that are normalized at $r_0$, assuming a prior background (in our case, the value from polynomial fitting in the reduction pipeline). An example is shown in Figure \ref{fig:n0_ngc5907} for the case presented in Section \ref{Sec:ngc5907}. The profiles are normalized at $r_0$ = 30{\arcsec} to have the same surface brightness that corresponds to a magnitude 0 star imaged by Dragonfly. Beyond 40{\arcsec}, individual profiles from might show flattening because of scattered light, or show steeping due to a biased background. We assign a normal distribution centered at $\hat{n_0}$ with a dispersion of $\delta \hat{n_0}$ as the new prior. The dispersion accounts for possible deviation in $n_0$ relative to $\hat{n_0}$, e.g., due to a change in the background. In a less crowded field, $n_0$ can also chosen to be fixed in the fitting.
    
    For the subsequent power law components of the PSF, our expectation on the extended PSF wing is that it will be flattened relative to the first (optimal) component by atmospheric or instrumental conditions. This is consistent with the profiles of isolated, luminous stars in Dragonfly images. Therefore, a shallower subsequent component is proposed and accepted if the likelihood is higher, otherwise the previous one extends through, which narrows down the prior space. In effect, the [$k+1$]-th power index $n_{k+1}$ is conditionally chained adopting a uniform prior ${U}[{\rm max}\{n_{\rm min},\, n_k-\Delta n\}, n_k-\delta n]$, where $n_k$ is the $k$-th power index. The two user-defined parameters, $\delta n$ and $\Delta n$, are set such that there exists a significant step $\delta n$ between each component while avoiding a strong discontinuity. In addition, a minimum slope $n_{\rm min}$ is set to avoid an over-shallow wing (which is time-consuming to render and could cause the computation to stall). The first transition radius $\log\theta_1$ follows a log-uniform prior ranging from $\log\theta_{\rm min}$ to $\log\theta_{\rm max}$. The subsequent transition radii follow log-uniform priors ranging from the previous radius to $\log\theta_{\rm max}$. We note, however, that a potential cutoff could exist, which is likely from background subtraction at a finite scale during the data reduction. Therefore, we include a cutoff option to mimic a possible drop. 
    
    For the sky modeling, by measuring the $3\sigma$-clipped statistics using the masked sky, one can get estimates of the mean and variance of the sky background, $\hat{\mu}$ and $\hat{\sigma}$. As noted before and illustrated in Figure \ref{fig:schematic}, these measurements are biased due to the scattered light, but they are useful as priors. We apply a truncated normal distribution for the mean sky background and a truncated log-normal distribution for its standard deviation, where each lobe on the right side of the mean is cut off. The polynomial coefficients follow uniform priors from 0 to $\hat{\sigma}$, assuming that the background variation is comparable to or under the level of its variance.
    
    Finally, we adopt a log-uniform prior for $f_{p}$, the fraction of light in the aureole part, from 1\% to 40\%. Because we adopt the normalization converted from the brightness $I_0$ rather than the total flux and mask the PSF core, the inner part of the PSF is largely decoupled from the aureole (i.e. with small covariance), which allows $f_{p}$ to be constrained in advance as well. We use parameters of the PSF cores from a maximum likelihood fitting on the median stacked PSF from unsaturated stars. This is also part of products produced by the \texttt{mrf}\footnote{\href{https://github.com/AstroJacobLi/mrf}{https://github.com/AstroJacobLi/mrf}} package (\citealt{2020PASP..132g4503V}). The \texttt{mrf} package carefully selects intermediate bright stars in a single image to generate a stacked non-parametric PSF model for the inner parts.
    
    One can increase the flexibility of the modeling by increasing the number of components included in the models to approximate an arbitrary curve. However, as the volume of prior space increases polynomially, the computational demand also dramatically increases. For Dragonfly, in most cases the extended PSF wing can be well modeled with 2-4 components out to 20{\arcmin}--30{\arcmin}, as shown in Section \ref{Sec:Case}.

\subsection{Computational Aspects} \label{Sec:realization}

We now turn to the computational details of the algorithm presented in the preceding sections. For concreteness, we will describe our approach by referring to the publicly-available package {\tt elderflower}, though of course the method described does not rely on our specific implementation. 

As has already been described, the central idea in our fitting pipeline is to render a 2D model of an area of the sky made up of a multitude of bright stars superposed on a slowly varying background sky model. Fitting on a pixel level is faced with the drawback that the computation is much slower than fitting a profile. The time cost depends on the cutout image size, the number of bright stars, the degree of freedom of the PSF model, and the volume of the prior space. 

Bright stars can be rendered in two ways: direct drawing in real space sampled by pixel response, or by convolution in the Fourier space using kernels generated from the model. These two approaches are identical only under the circumstance that the kernel size is larger than or at least comparable to the image size. Using a small kernel could leave artifacts on the final image. These artifacts can be non-negligible, especially for very bright stars whose scattered light pervades the entire image. On the other hand, the kernel size should not be too large; otherwise, the convolution step becomes too slow. The safer method is rendering in real space, but it quickly becomes slow as the image size increases since the time cost is equal for all sources, making it impractical to draw all the sources in real space.
Given the advantages and disadvantages of these two rendering approaches, a straightforward strategy is to split the model stars according to their brightness, as in Section \ref{Sec:general_consideration}; VB stars are drawn in real space to avoid truncation effect, and MB stars are drawn by convolution in the Fourier space for efficiency. 

We use the image simulation tool \texttt{GALSIM} (\citealt{2015A&C....10..121R}) as the engine to draw stars by convolution. \texttt{GALSIM} provides a software library whose bulk calculations are carried out in C++ with excellent performance in both efficiency and precision. The kernel is generated by the PSF model and interpolated within \texttt{GALSIM}. A \texttt{GALSIM} object uses the total flux as normalization. Therefore, the normalization is given by the integration of the PSF scaled by $\hat{I}_{*,i}$ (Eq. \ref{Eq:I_int}). To improve efficiency, we assign an adaptive kernel size according to the contrast of the PSF shape, where a PSF descending more gently has a larger kernel size. Because \texttt{GALSIM} renders objects independently, model stars can be drawn in parallel. Parallel computing is enabled when the computation becomes lengthy or under the circumstance that the field is crowded, otherwise we switch back to serialized computation to avoid overheads.

For posterior sampling, we use the dynamic nested sampling codes \texttt{dynesty} \citep{2020MNRAS.493.3132S}. We adopt the uniform sampling method, multiple ellipsoid bounds and a mixture of 80\% posterior and 20\% evidence. The prior and stopping criteria are also proposed in parallel. The time required to fulfill the stopping criteria depends on the number of `live points' used, the model validity, and the volume of prior space. Because in typical cases (specifically, in the absence of large-scale contamination such as Galactic cirrus) the posterior of the PSF+sky is well defined with single modes, we use a small number of live points as default ($N_{\rm live}=\texttt{ndim}\cdot10$, where \texttt{ndim} is the number of parameters). Increasing the number of live points used in the sampling would improve the resolution (`smoothness') of posteriors but also increase the computation time.

\section{The Dragonfly PSF: Application to the M44 Field}  \label{Sec:DF_PSF} 

In this section, we apply \texttt{elderflower} to an open cluster, Messier 44 (M44), observed by Dragonfly to test our methodology of PSF modeling. We present the optimal PSF model retrieved by \texttt{elderflower} and compare it with SDSS and wide-angle PSFs of other telescopes in the literature. The primary purpose here is to demonstrate the efficacy of the methodology proposed in Section \ref{Sec:Methodology}, which works under the condition that scattered light is significant and non-negligible, such as for the challenging dataset described here.

\subsection{Modeling of Wide-angle PSF in the M44 Field} \label{Sec:modeling_m44}

    The visible open cluster M44, also known as the `Beehive' cluster or the Praesepe cluster, is one of the nearest star clusters. It extends $\sim$ 1.5{\degree} in the sky and contains around 900 star members (\citealt{2007AJ....134.2340K}). Bright stars in M44 are up to around 6 mag, with over 200 stars in the field brighter than 13 mag and around 60 stars brighter than 10 mag.
    
\begin{figure}[!htbp]
    \centering
      \resizebox{\hsize}{!}{\includegraphics{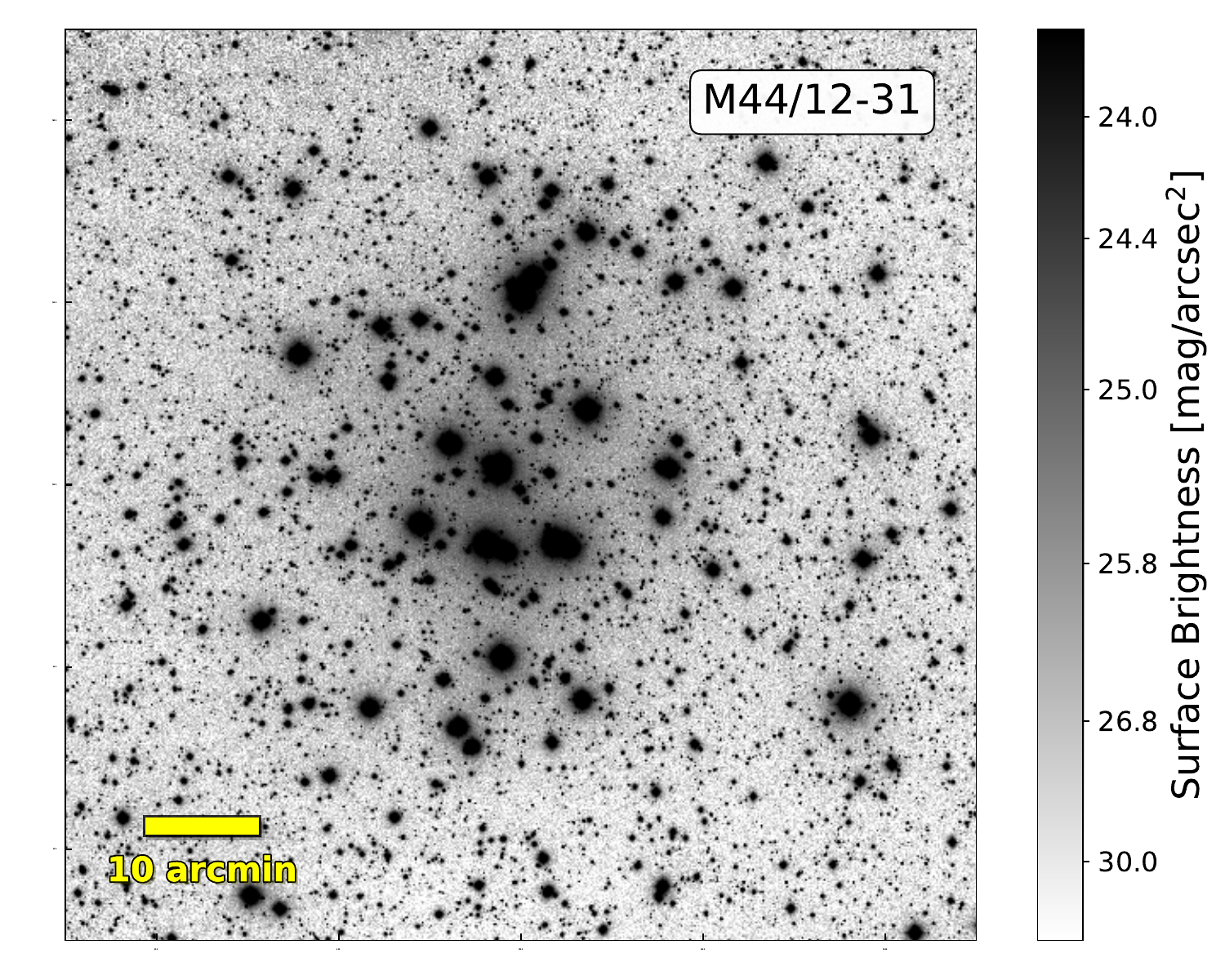}}
      \caption{1.6{\degree}$\times$1.6{\degree} g-band image of the open cluster M44 taken by Dragonfly on 2019 December 31, after lens cleaning on 2019 December 30. The image scale is stretched to enhance low surface brightness levels.}
    \label{fig:m44-1231_image}
    \end{figure}

    The images of M44 were taken by Dragonfly on two nights in 2019 December, before and after its seasonal lens cleaning. The image on the first night is slightly deeper than the second night. Unlike the data products produced by the standard Dragonfly reduction pipeline, the original pixel scale of 2.85{\arcsec}/pix is retained. The central 1.6{\degree}$\times$1.6{\degree} $g$-band cutouts of the data taken on the second night are shown in Figure \ref{fig:m44-1231_image}. The scattered light is prominent in the central 1{\degree}$\times$1{\degree} region of the cluster.{ The central 0.8{\degree}$\times$0.8{\degree} region is used for the PSF modeling.} Given that the M44 field is an extremely crowded field filled with bright stars, it serves as a good test of whether our wide-angle PSF modeling is effective in such conditions where the extended PSF wings are totally non-negligible. We take advantage of the absence of gaseous structure in M44 (unlike, e.g., M45), which could affect the PSF modeling. The motivation of taking two night data in a row is to quantify the difference of the wide-angle PSF between the two nights, which is expanded in Section \ref{Sec:m44_dust}. Below we proceed with our modeling with the second night data of M44.
    
    \begin{figure*}[!htbp]
    \centering
      \resizebox{\hsize}{!}{\includegraphics{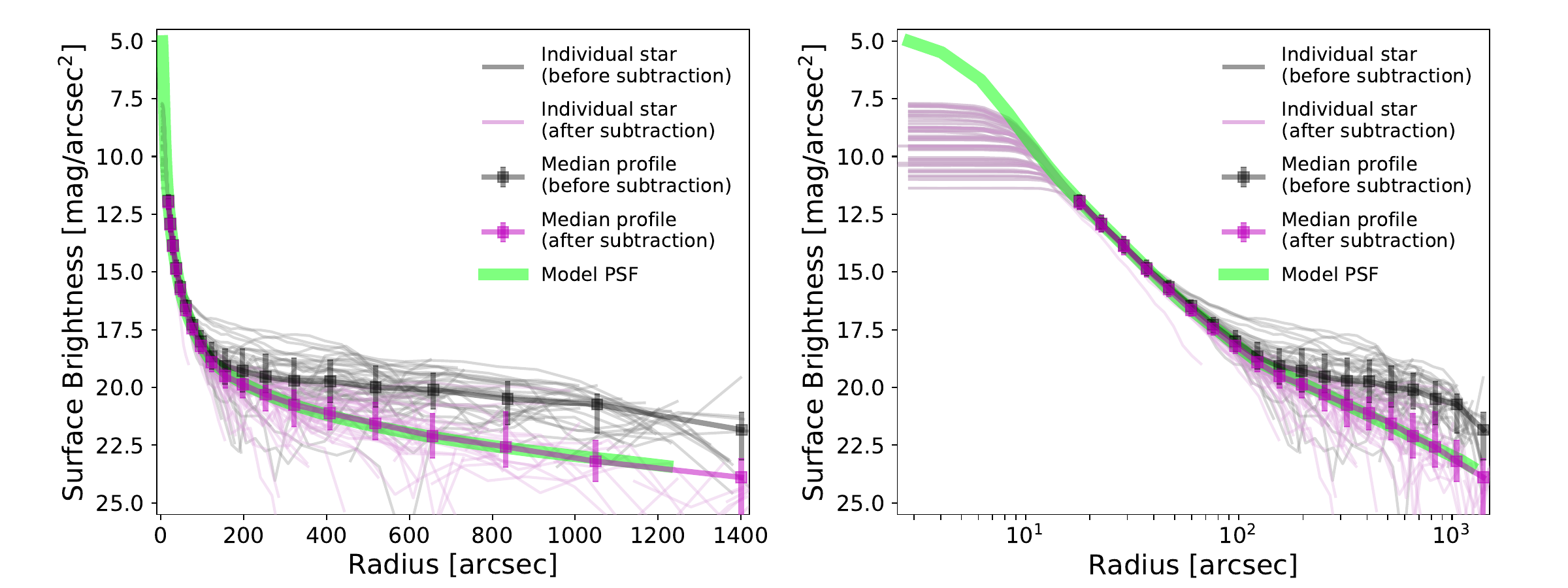}}
      \caption{Model PSF derived from the second night (2019-12-31) data of M44 (green curves) and profiles of individual stars brighter than 10 mag, before (black) and after (magenta) subtraction of scattered light using the model PSF. The thicker curves in square symbols show the median stacking of profiles with the corresponding color. The left/right panel displays the surface brightness profiles on a linear/log scale. The model PSF is normalized to have its total flux equivalent to a magnitude 0 star integrated out to its maximum range. Individual stellar profiles are scaled by the mean surface brightness between 20{\arcsec} and 40{\arcsec}.}
    \label{fig:psf_m44}
    \end{figure*}
    
    Because of the large number of bright stars, several special compromises are required to allow the fits to be obtained in a reasonable time on a modest hardware. We only include stars brighter than 12.5 mag in the model (MB+VB stars) and treat those below 10 mag as VB stars. We adopt a three-component model with a constant background. No cutoff is applied given the large number of bright stars. The scaling radius $r_0$ is 10 pix, which corresponds to $\sim$30\arcsec. We have checked that no saturation occurs outside of this range. The PSF is modeled out to 20{\arcmin}. Aperture masks with radii of $\sim$35\arcsec (corresponding to 12 pix) are used to mask the stellar cores. The fraction of the power law component $f_{p}$ is fixed to be the value obtained from the stack. Because extended wings of VB stars dominate the scattered light in the field whose normalizations are determined from their brightness at specific radii, slight variation in $f_{p}$ would not affect the result. The final PSF model is displayed as the green curve in each panel of Figure \ref{fig:psf_m44} (left: linear scale; right: log scale), {described by the parameter set ${\bf\Theta}:\{f_{p},\, \alpha,\, \gamma,\, n_0,\, n_1,\, n_2,\, \log\theta_0,\, \log\theta_1,\, \log\theta_2\}=\{0.3,\, 6.7,\, 6.1,\, 3.62,\, 2.90,\, 1.89,\, 0.7,\, 1.73,\, 2.1\}$ ($\gamma$ and $\theta_k$ in arcsec), with $1\sigma$ uncertainties of the fitted aureole parameters to be $<$ 3\%.} It is normalized in a way that the total flux integrated out to its maximum range is equivalent to a magnitude 0 star. 
    
    As a test, we extract the normalized surface brightness profile of each individual star brighter than 10 mag before and after scattered light subtraction. {The subtraction leaves out the light of the individual star being profiled.} The background values are measured with a 2{\arcmin} wide annulus at 20{\arcmin}. The extracted stellar profiles are normalized with the mean surface brightness between 20{\arcsec} and 40{\arcsec}. These individual profiles are shown as magenta and black curves, with and without scattered light subtraction, respectively. The thicker magenta/black curve represents the median stacking of individual profiles of the corresponding color. The magenta stacking profile matches well with the model PSF, which indicates the self-consistency of our method, while the black stacking profile is significantly flattened and then drops at large radii. The individual stellar profiles in the presence of scattered light also have a larger dispersion: most are flattened by the scattered light from bright stars, while a small fraction of them show a rapid falloff due to a biased background. 
    
    Figure \ref{fig:psf_m44} demonstrates why stacking stars is not preferred for the reconstruction of an accurate and unbiased PSF model: contaminating scattered light is impacting the stellar profiles, and the embedded systematics are simply propagated at higher S/N levels in a stack.
    In summary, this `leave-one-out' experiment illustrates that: (1) our method works under extreme conditions such as in M44 where scattered light pervades the field; and (2) the classical method, which ignores the scattered light, can lead to significant bias in the wide-angle PSF when the scattered light in the field is prominent (as described in Section \ref{Sec:challenge}).
    
    {An alternative approach to testing our methodology is to apply it to a simulated image where the underlying PSF is known. An example of such a test is presented in Appendix A, and the conclusion of this experiment is that the output PSF model is consistent with the input PSF, as expected. See Appendix \ref{Sec:fit_mock} for details.}

\subsection{Comparison with SDSS} \label{Sec:sdss_check}
    
    \begin{figure}[!htbp]
    \centering
      \resizebox{\hsize}{!}{\includegraphics{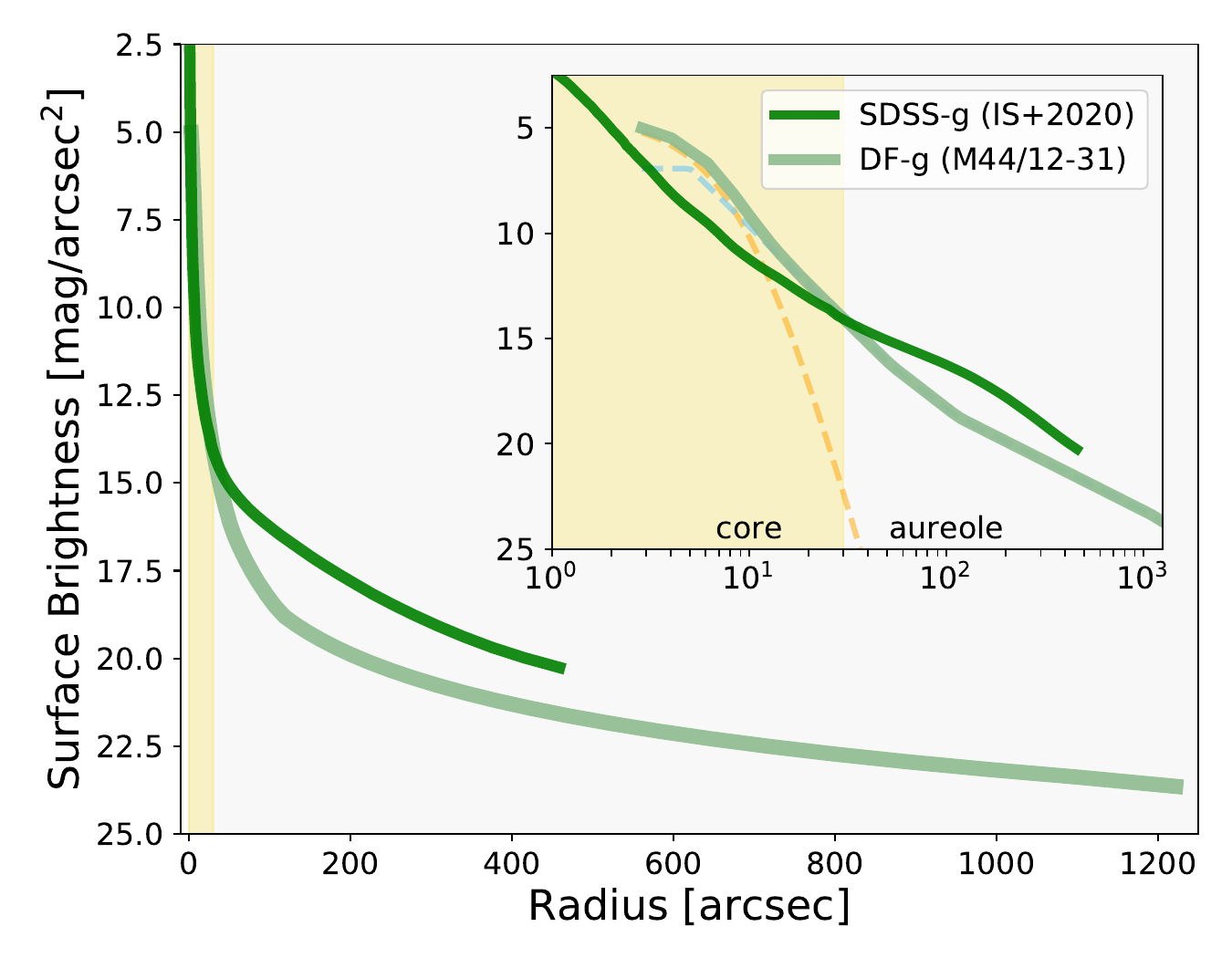}}
      \caption{Model PSF derived from the second night (2019-12-31) of g-band M44 data compared with SDSS g-band PSF, displayed in log scale (major panel) and linear scale (inserted panel). The Moffat component and the multi-power law component are shown in orange and blue dashed curves. The SDSS PSF extends out to 8{\arcmin} and is built with stacking techniques (\citealt{2020MNRAS.491.5317I}). The Dragonfly and SDSS PSFs are normalized to have a total flux of a magnitude 0 star integrated out to its maximum range.}
    \label{fig:psf_m44_sdss}
    \end{figure}
    
    We compare the Dragonfly model PSF with the wide-angle PSF of SDSS in Figure \ref{fig:psf_m44_sdss}. The SDSS PSF is measured using the stacking technique by \cite{2020MNRAS.491.5317I} out to 8{\arcmin}.
    The PSFs are scaled to have the same total flux integrated to their maximum range corresponding to a magnitude 0 star. Compared with SDSS, Dragonfly's resolution is poor, so it is not good at suppressing light in the core region where the Moffat core dominates. However, the suppression of scattered light in the wide-angle PSF of Dragonfly is remarkably superior to that of SDSS. Note that the SDSS PSFs steepens beyond 3{\arcmin}--4{\arcmin}, possibly due to the finite chip size of SDSS used for the measurements.  

    \begin{figure*}[!htbp]
    \centering
      \resizebox{\hsize}{!}{\includegraphics{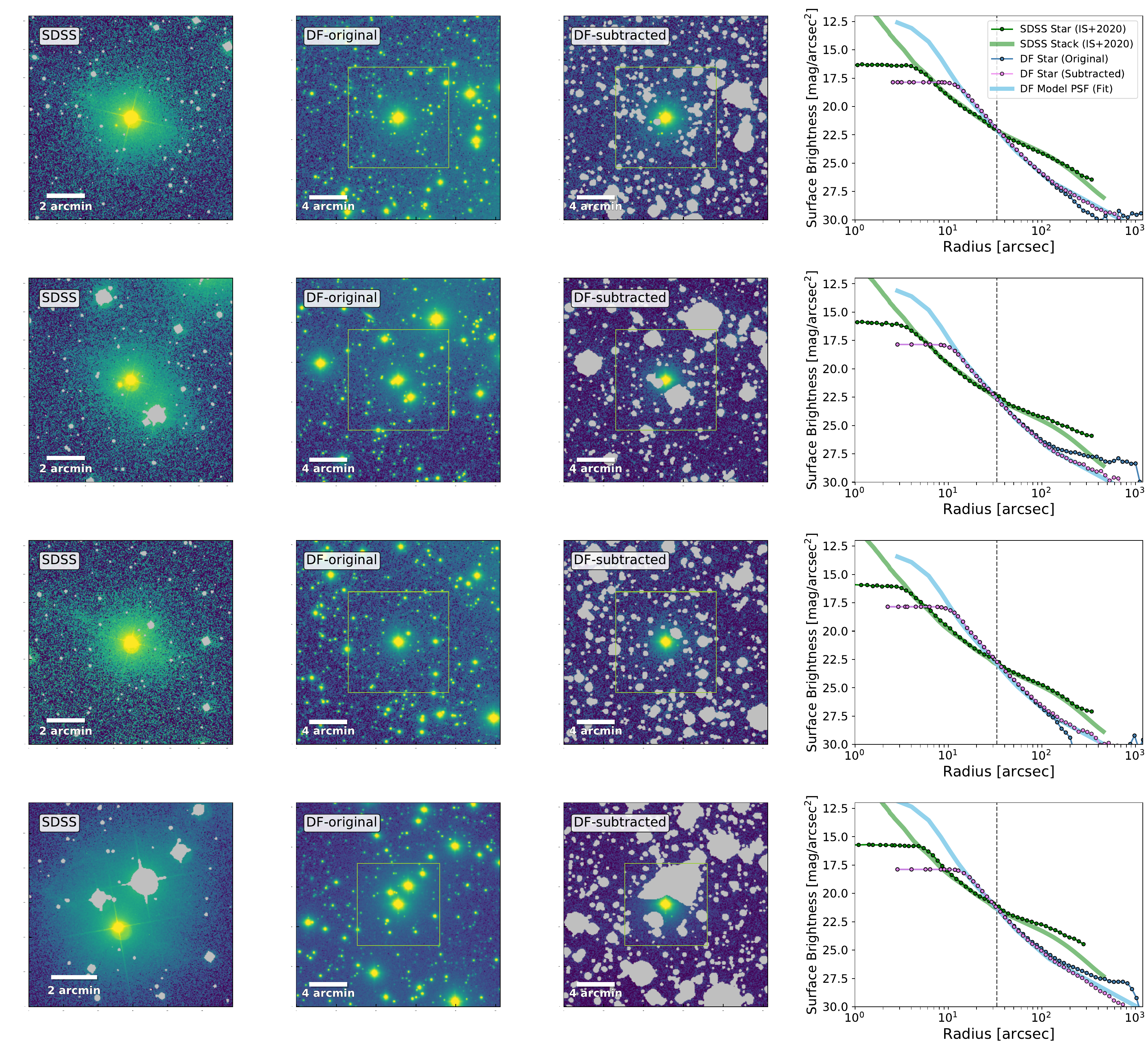}}
      \caption{A comparison of images and surface brightness profiles for several individual stars in the field of M44, imaged by SDSS and Dragonfly. {\em 1st column:} images of stars in SDSS. {\em 2nd column:} images of stars in Dragonfly, without scattered light subtraction.  {\em 3rd column:} images of stars in Dragonfly, with scattered light subtracted. In the image stamps, masked regions are shown in gray, while masks in the 2nd column are not shown for display purpose. The SDSS footprints in Dragonfly images are indicated by yellow squares. {\em 4th column:} extracted profiles from SDSS (green symbols) and Dragonfly (blue/magenta symbols). The SDSS PSF (obtained from stacking; \citealt{2020MNRAS.491.5317I}) and the Dragonfly model PSF are normalized to have their total flux equal to the magnitude of the target star and shown as the thick green and blue curve, respectively, in each panel. The black dotted lines indicate where Dragonfly and SDSS PSFs intersect. The relative offsets between the extracted profiles of SDSS and Dragonfly are similar. The extracted profiles with scattered light subtraction (in magenta) also match the model profiles well. See text for details.}
    \label{fig:check_sdss_star}
    \end{figure*}
    
      Given the large difference in the pixel scale and the PSF shape between Dragonfly and SDSS, a reasonable comparison between them also needs to be based on a reliable normalization. Therefore, we perform a consistency test on this normalization by directly measuring surface brightness profiles of the same stars. These are chosen to be away from the cluster center in the M44 field, so that they are less affected by scattered light. The profiles are extracted in the same way as in the previous section. Nearby sources are masked in the measurement.
      
      Cutout images of stars from the SDSS and Dragonfly are shown as stamps in the first to third columns of Figure \ref{fig:check_sdss_star}. The first column shows SDSS images of the selected stars. The SDSS images are retrieved from the DR12 Science Archive Server with a pixel scale of 0.5 {\arcsec}/pixel and combined into a 0.2{\degree}$\times$ 0.2{\degree} mosaic with 35-45 frames centering at each star\footnote{We use the \texttt{frame} data model products of SDSS for the validation. The \texttt{frame} data model has a global sky subtraction where a large-scale sky model is calculated with spline fitting using the full set of fields for each run, which is limited by the chip size in one direction but not in the other. We refer the readers to \cite{2011ApJS..193...29A},  \cite{2011AJ....142...31B}, and the documentation of the data model for the processing details of the used SDSS data.}. The second and third columns show the same stars in Dragonfly images, with and without scattered light subtraction (see the previous section), respectively.
      
      The fourth column shows the extracted profiles in green (SDSS) and blue/magenta (Dragonfly; without/with scattered light subtraction). For comparison, the stacked PSF and the fitted Dragonfly PSF model normalized to the corrected magnitude $m_{corr}$ are shown as thick curves in green and blue. The magnitudes (corrected from the Pan-STARRS catalog) of these stars away from the cluster center are less affected by the scattered light. Although the cores are saturated and the outskirts of the PSF present differences, {the relative offsets between the extracted SDSS and Dragonfly profiles are consistent, with similar crossover at radii around 30-40{\arcsec}. The extracted Dragonfly stellar profiles with scattered light subtracted also match the fitted PSF model well out to large radii. Therefore, we can conclude that the normalization between SDSS and our model PSF is consistent.} Note at large radii the measured profiles before scattered light subtraction suffer from bias, appearing either flattened due to scattered light, or steepened due to a biased background.

\subsection{Comparison with Wide-Angle PSFs in the Literature}
    
    \begin{figure*}[!htbp]
    \centering
      \resizebox{\hsize}{!}{\includegraphics{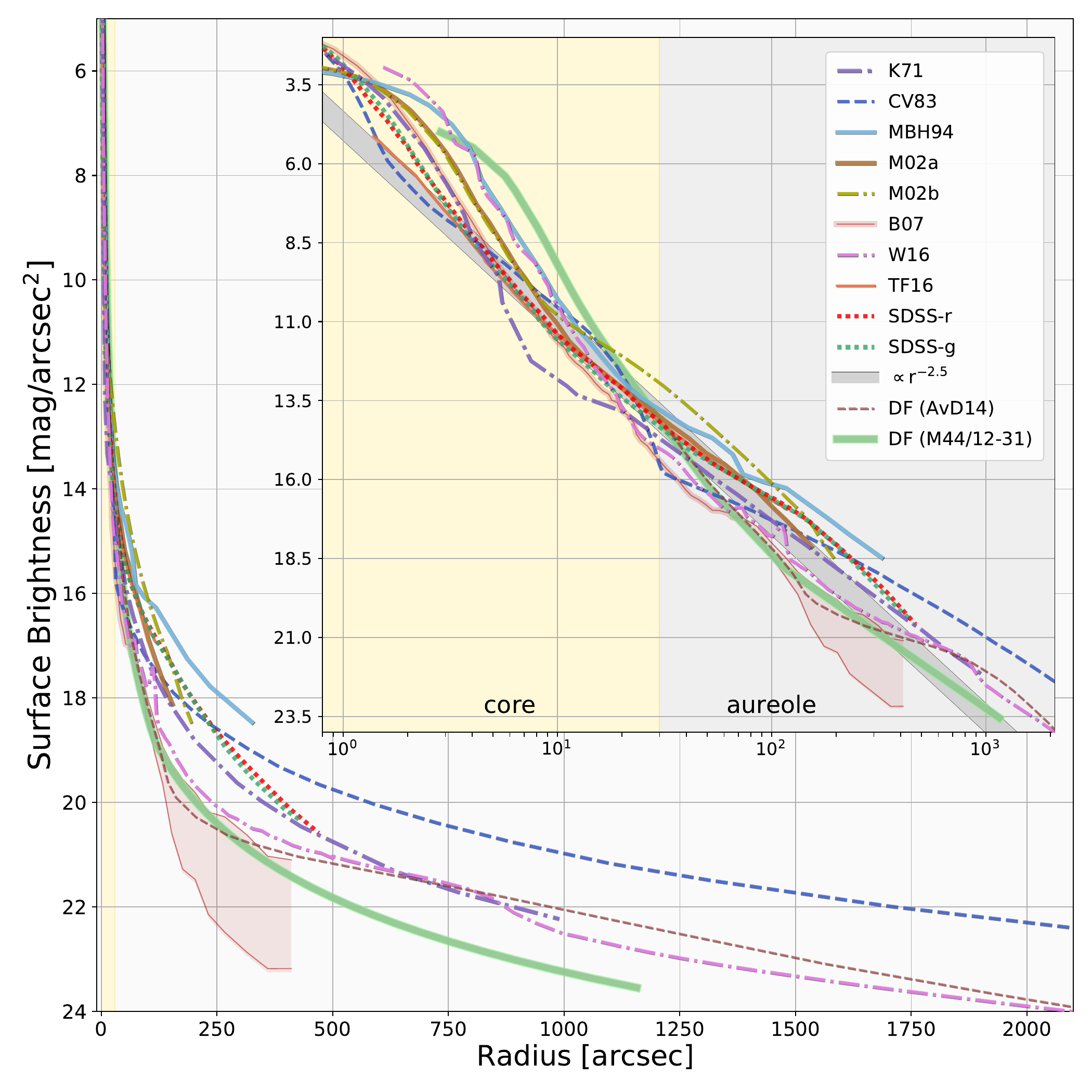}}
      \caption{Surface brightness profiles of a magnitude 0 star imaged by Dragonfly, SDSS, and {seven other telescopes on a linear (major) and log (minor) scale. The SDSS PSFs are obtained through stacking techniques (\citealt{2020MNRAS.491.5317I}).} The Dragonfly PSF modeled from the second night g-band data of M44 is displayed as the green solid line. The outer portion of an early r-band measurement of the Dragonfly PSF using Vega is also overplotted (AvD14). 
      An $r^{-2.5}$ power law is displayed as the gray band. See text details for the normalization of profiles. {\em References}: K71: \cite{1971PASP...83..199K}; CV83: \cite{1983ApJS...52..465C}; M02a, M02b: \cite{2002A&A...384..763M}; MBH94: \cite{1994AJ....108.1191M}; B07: \cite{2007ApJ...666..663B}; AvD14: \cite{2014PASP..126...55A}; TF16: \cite{2016ApJ...823..123T}; W16: \cite{Watkins2016}.
      }
    \label{fig:psf_literature}
    \end{figure*}
    
    Figure \ref{fig:psf_literature} compares the Dragonfly wide-angle PSF derived from the M44 field to those from a broader range of facilities. Most of the other wide-angle PSF measurements taken from the literature were compiled by \cite{2014A&A...567A..97S}, to which the reader is referred for detailed information about the measurements, telescopes, observational conditions, etc. 
    
    In brief, the PSF labeled $\rm PSF_{K71}$ is from the work of \cite{1971PASP...83..199K} composed of several measurements out to 5\degree; $\rm PSF_{CV83}$ is from \cite{1983ApJS...52..465C}, which reaches the widest radius (r = 90\degree); $\rm PSF_{MBH94}$ is from \cite{1994AJ....108.1191M} who detected the extended faint halo of NGC 5907; $\rm PSF_{M02a}$ and $\rm PSF_{M02b}$ are from \cite{2002A&A...384..763M}, who first studied the temporal variation of the wide-angle PSF, with measurements separated by three months.
    We follow the normalization of \cite{2014A&A...567A..97S} for $\rm PSF_{CV1}$, $\rm PSF_{K71}$, $\rm PSF_{M02a}$, $\rm PSF_{M02b}$, and $\rm PSF_{MBH94}$.
    
    {$\rm PSF_{B07}$ represents a rendering of the PSF of the 2.5m du Pont telescope at Las Campanas Observatory (\citealt{2007ApJ...666..663B}), with the shaded area enclosing the measurements of an off-axis bright star (their 2000 profile, see Figure 1 of \cite{2007ApJ...666..663B}). Besides a bump between 60{\arcsec} and 100{\arcsec} likely caused by reflection between the CCD and the dewar window/filter, $\rm PSF_{B07}$ follows a power law with a power index of $n=2.5-3$. We shift $\rm PSF_{B07}$ by 0.75 mag to match its integral of flux with the SDSS PSFs. It is worth noting that the outer portion of PSF$_{\rm B07}$ is steeper than the PSF of other reflecting telescopes.}
    $\rm PSF_{W16}$ represents the V-band PSF of the 0.9m Burrell Schmidt telescope on Kitt Peak (\citealt{Watkins2016}), the telescope that has produced the well-known detection of diffuse intracluster light in Virgo cluster (\citealt{2005ApJ...631L..41M}).
    It is an updated version of the profile in \cite{2009PASP..121.1267S} with the inner region scaled properly with the wing.
    The extended wing follows a power law of $r^{-2.4}$ (\citealt{2009PASP..121.1267S}), which is very similar to SDSS PSFs ($r^{-2.5}$). We slightly shift $\rm PSF_{W16}$ to match the 1D integral with other profiles. $\rm PSF_{TF16}$ represents the PSF of the Gran Telescopio de Canarias telescope (\citealt{2016ApJ...823..123T}) used for deep imaging on ultra diffuse galaxies. We shift $\rm PSF_{TF16}$ to match $\rm PSF_{S09}$ in their overlapping range since their slopes are also similar.
    
    The Dragonfly and SDSS PSFs are scaled to 0 mag, as in Section \ref{Sec:sdss_check}, to match the representation of the other profiles.
    The Dragonfly PSF has the most suppressed extended wing when compared with other PSF measurements. Dragonfly is the only telescope using all-refractive optics, which suggests an instrumental origin for the differences in the primary component of the wide-angle PSF. Within $\sim$2{\arcmin}, the Dragonfly wide-angle PSF approximates the $r^{-3}$ law expected from pure aperture diffraction\footnote{Light diffracted by a circular aperture, i.e., the Airy disk, asymptotically falls off as $r^{-3}$ at very large angles, although the exact slope depends on the scale over which the PSF is averaged. In practice, the Fresnel rings are smeared out by seeing, the finite bandpass and, in Dragonfly, the large pixel size. Other instrumental/atmospheric factors could contribute to the actual radial dependence, making it challenging to explain the origin of the exact slope of PSF.}. 
    Note that this figure shows the Dragonfly PSF from the second night data of M44 imaging, taken after lens cleaning. After cleaning, at very large angles, Dragonfly’s PSF has a power index of $n=1.9$, which is close to the $r^{-2}$ law found at the outer range of $\rm PSF_{K71}$ in \cite{1971PASP...83..199K}. The origin of this $r^{-2}$ component is still unclear, with one possible source being scattering from atmospheric cirri or aerosols in the atmosphere (\citealt{2013JGRD..118.5679D}). 
    
    We also overplot an early measurement of the Dragonfly PSF using Vega (\citealt{2014PASP..126...55A}) as the light pink curve in Figure \ref{fig:psf_literature}. The PSF was built by stitching several sub-profiles with a range of integration times, extending out to around 1 {\degree}. We plot only the wide-angle portion of the PSF and rescale it to match the first component of the new model, since it has a slope similar to the latter out to around 2{\arcmin}. The early measurement becomes slightly shallower than the upper boundary of the new model, and then rapidly falls off beyond 30{\arcmin}. It is not clear what leads to the slope difference in the two measurements, e.g., whether the flattening in the early measurement is caused by dust or whether the drop-off persists in the new camera models. 
    
    Finally, we note that the current version of the Dragonfly data reduction pipeline removes background structures on scales larger than $\sim$45{\arcmin} (see Section \ref{Sec:Discussion}), which limits the modeling range of publicly released Dragonfly observations. We will perform further investigations of the PSF beyond degree scales in a future paper after improvements to the pipeline and further refinement of our PSF modeling method.

\section{Additional Case Studies}  \label{Sec:Case} 

In this section we present further examples of wide-angle PSF modeling using data from the Dragonfly telescope. The fittings are run through \texttt{elderflower} to obtain the optimal wide-angle PSF models. These case studies show the PSF obtained in a less extreme scattered light situation than exemplified by the M44 observations, and touch on the importance of lens cleanliness for optimal wide-angle PSF light suppression. The latter will be explored in greater depth in a follow-up paper, together with the subject of variability and chromatic dependence of the Dragonfly wide-angle PSF.

\subsection{NGC 5907 Field: A Bright Star Trio}  \label{Sec:ngc5907} 

\begin{figure*}[!htbp]
\centering
  \resizebox{0.92\hsize}{!}{\includegraphics{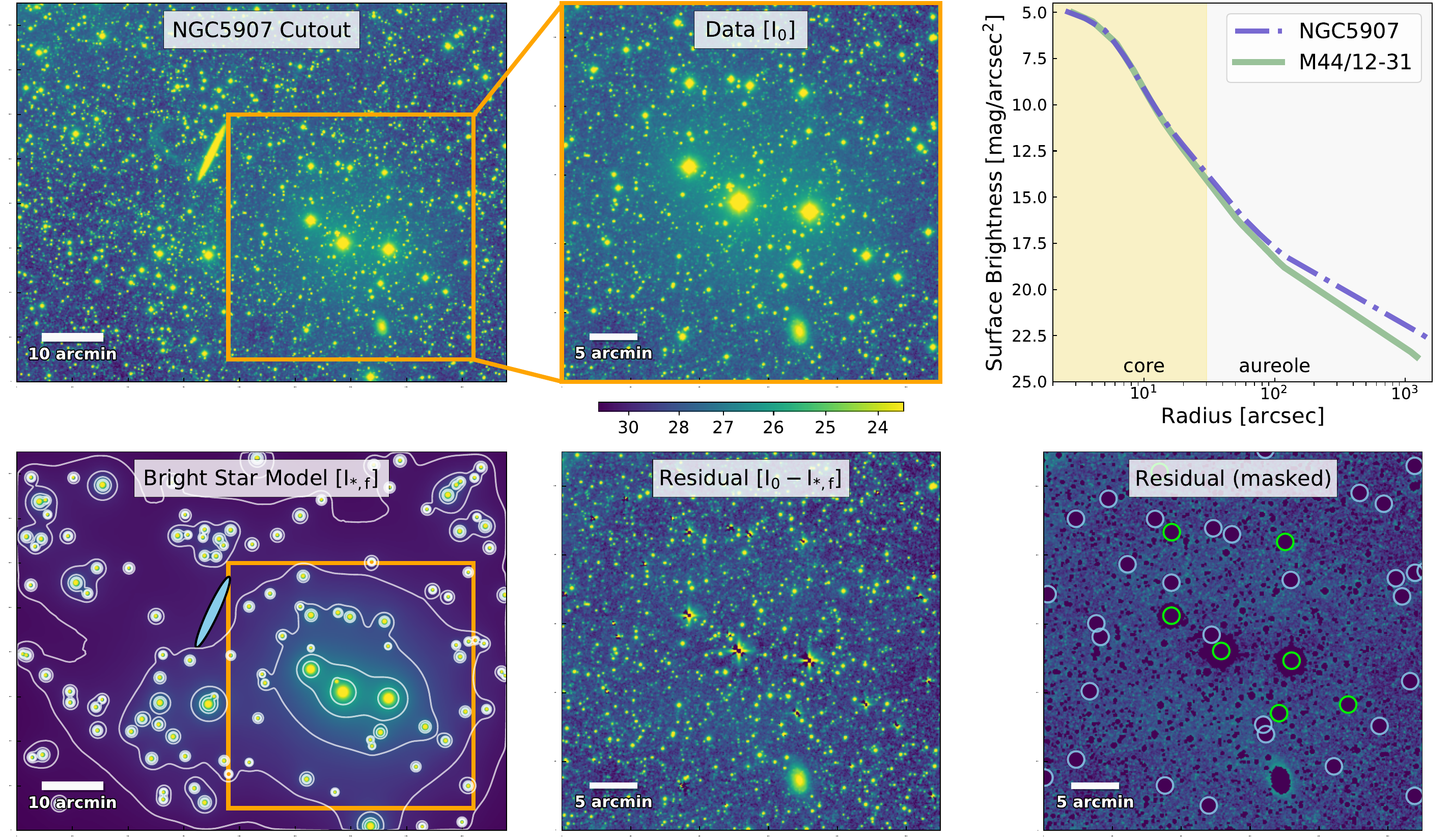}}
  \caption{{\em Top left}: $1.5\degree \times 1.2\degree$ cutout of NGC 5907 imaged by Dragonfly. {\em Top middle}: the 50$\arcmin\times$50$\arcmin$ zoom-in region to be fitted. {\em Top right}: {model PSF derived from the region. For comparison, the PSF from the second night of M44 imaging is also plotted.} {\em Bottom left}: Reconstructed bright stars ($<$15 mag) in the field based on the fitted model. {Contours indicate surface brightness levels from 30 to 26 mag/arcsec$^2$ at an interval of 1 mag/arcsec$^2$.} The edge-on galaxy NGC 5907 is illustrated as the blue ellipse. {\em Bottom middle}: the data image (panel above) after subtraction of the reconstructed bright stars (panel to left). {\em Bottom right}: as in the bottom middle panel but with core regions and extended sources masked. Green and blue circles represent VB and MB stars in the model, respectively. The image scales are the same and stretched to enhance low surface brightness levels.}
\label{fig:fit_ngc5907}
\end{figure*}

NGC 5907 is a nearby edge-on galaxy that has been imaged by Dragonfly down to low surface brightness levels for the study of its tidal features (\citealt{2019ApJ...883L..32V}). To the southwest of the galaxy (0.5 deg apart), there are three bright stars ($\sim$7-8 mag in g-band) whose joint effect in scattered light is considerable. The combined scattered light from the wide-angle PSF extends at least 20{\arcmin} away, impacting a wide area of sky around them, and potentially contaminates the measurement of NGC 5907, even though the galaxy appears to be at a `safe' distance from these stars. 
On the other hand, in terms of their brightness and proximity to the target galaxy, these stars are ideal targets from which one is able to extract a good representation of the local PSF (i.e., against spatial variation) for the target galaxy out to a large radius. This is challenging with classical methods since photometry on any single star of the three would be biased by the other two. As a result, this portion of the NGC 5907 field serves as a nice illustration of our approach of wide-angle PSF modeling.

The full image of NGC 5907 obtained by Dragonfly has a field of view of $\sim3\degree \times 4\degree$ and a pixel scale of 2.5\arcsec/pix. {NGC 5907 is one of the targets of the Dragonfly Edge-on Galaxy Survey (DEGS) (Gilhuly et al., submitted)}. In total, 618 frames were used to make the final stack, equivalent to an exposure of $\sim$2 hours with Dragonfly. The top left panel of Figure \ref{fig:fit_ngc5907} displays the central $1.5\degree \times 1.2\degree$ of the Dragonfly image. We cut out a region of $50\arcmin \times 50\arcmin$ around the central bright star. The cutout region used for the PSF fitting is shown in the top middle panel of Figure \ref{fig:fit_ngc5907}. 


A four-component aureole model was adopted for the fitting. The central 1{\arcmin} regions of bright stars were masked and the PSF was modeled out to 25{\arcmin}. Fitting results are presented in Figure \ref{fig:fit_ngc5907}: 
The bottom left panel shows the background reconstructed from PSF wings of bright stars. As shown by the contours, scattered light from extended PSF wings pervades the field. {The top right panel shows the fitted PSF model from the 50$\arcmin\times$50$\arcmin$ cutout. The PSF derived from M44 imaging in Section \ref{Sec:modeling_m44} is also plotted for comparison. The PSF from the NGC 5907 field has a shallower outer wing ($n=1.6$) than that of M44 ($n=1.9$).} The bottom middle panel displays the residual where the reconstructed scattered light from bright stars has been subtracted from the original image. The cores of bright stars are not recovered well; this is expected because our core PSF model is crudely built without consideration of any high-order effects such as spikes and tracking errors. However, the large-scale scattered light that we are trying to model has been largely eliminated, without a visually significant spatial pattern associated with bright stars appearing in the residual. In the south of the subtracted image a disk galaxy is preserved, since extended sources are not included in the modeling. Finally, the bottom right panel displays the same image after masking stars -- the field appears to be fairly flat without significant underlying patterns (Galactic cirrus, etc.) once the scattered stellar light is eliminated. 

It is interesting to consider whether failing to modeling the wide-angle PSF would have had a significant impact on the measured stellar halo of a galaxy placed in this field. This is explored with a photometric test on low surface brightness galaxies (LSBGs) in Appendix \ref{Sec:mock_test}, the conclusion of which is that modeling the background scattered light is essential for reliable profile estimation.

\subsection{Test on Ultrawide Fields Observed in a Sequence} \label{Sec:uw_test}

{Another interesting test is to run the wide-angle PSF modeling on data taken on nearby but different fields imaged on the same night but at different times to determine whether their wide-angle PSFs are similar.}

\begin{figure*}[!htbp]
\centering
  \resizebox{\hsize}{!}{\includegraphics{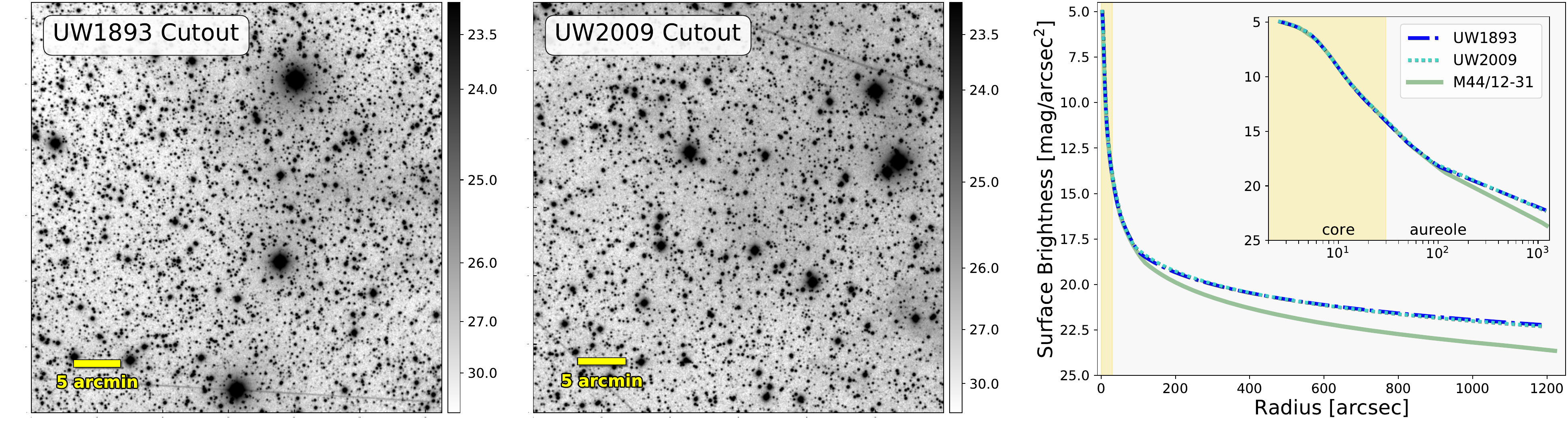}}
  \caption{{\em Left}: the used $50\arcmin \times 50\arcmin$ cutout in the UW1893 field. {\em Middle}: the used $50\arcmin \times 50\arcmin$ cutout in the UW2009 field.  {\em Right}: wide-angle PSF models fitted from two UW fields (UW1893: blue dash-dotted; UW2009: cyan dotted) observed within a small time frame ($\sim$1 hour) on a linear (major) / log (minor) scale. The surface brightness profiles are normalized to a magnitude 0 star. The PSF derived from the second night of M44 data is also displayed for reference.}
\label{fig:psf_uw}
\end{figure*}

{To undertake our test, we selected two fields from the Dragonfly Ultra Wide (UW) Survey
\footnote{{The Dragonfly UW survey is a relatively shallow (by Dragonfly standards) low surface brightness photometric survey complementary to SDSS. The survey has a typical 1$\sigma$ depth of 30 mag/arcsec$^2$ on $1\arcmin \times 1\arcmin$ scales in g-band with a pixel scale of 2.5\arcsec/pix. The survey strategy, science goals, and data assessment will be presented in a future paper. For details on the data reduction and qualities, we refer the readers to papers describing the Dragonfly Wide Field Survey, presented in \cite{2020ApJ...894..119D} and \cite{2021ApJ...909...74M}.}}, UW1893 and UW2009, out of the current dataset as the test fields here. The exposures of UW1893 and UW2009 were taken on the same night (2021 June 9) within a short time span (separated by around one hour). The two fields are less affected by large-scale Galactic cirrus. According to the weather records on the night, the weather conditions were clear throughout the night, with good seeing and low humidity.}

{For each field we cut out a $50\arcmin \times 50\arcmin$ region that included a few bright stars (7$\sim$9 mag in g-band) to run the modeling of the wide-field PSF. The two regions have approximately the same distance to the field center. The cutouts are displayed in the first and second panels of Figure \ref{fig:psf_uw}.}
{A three-component aureole model was adopted for the fitting, with the PSF being modeled out to 20{\arcmin}. The central 1{\arcmin} regions of bright stars were masked. The derived PSF models from the two fields are shown in the right panel of Figure \ref{fig:psf_uw}. For comparison, we also plot the PSF derived from the M44 field in Section \ref{Sec:modeling_m44}. The two derived PSFs are consistent, with  similar slopes in their power-law components: $n=3.6-3.7$ within 50{\arcsec}, $n=2.8-3$ between 50{\arcsec} and 90{\arcsec}, and $n=1.5-1.6$ beyond 90\arcsec. Their outer wings are shallower than that of the M44 field ($n=1.9$). Such consistency indicates that the `instant' PSFs during the short time period of observation for these two specific fields do not present significant changes and serve as further support to the efficacy of our method.}

{We have run such tests on several other fields taken consecutively, with a good number of them showing similar consistency. However, in some instances we did find some difference, principally in the outermost component. It is unclear whether relevant atmospheric conditions change during the hour-long interval, though this is certainly conceivable, given the known variation in the properties of the relevant aerosols, which are likely contributing sources for the wide-angle PSF (\citealt{2013JGRD..118.5679D}). The properties of aerosols can exhibit significant variability hourly, and even on sub-hourly timescales (e.g., \citealt{STANIER20043275}). An in-depth investigation of the temporal variability of the wide-angle PSF is beyond the scope of this paper and will be explored in a future work.}

{We also note that many of our fields present noticeable Galactic cirrus, which might bias the background and/or the estimate of the outer wings of the wide-angle PSF (see Section \ref{Sec:Discussion}). The contamination from the Galactic cirrus emission is being investigated and will be the subject of a subsequent paper.}

\subsection{Degradation of the Wide-angle PSF from Dust Accumulation on Lenses}  \label{Sec:m44_dust} 



We foreshadowed the {possible} impact of lens cleaning on the wide-angle PSF in Section \ref{Sec:DF_PSF}. Here we proceed with more discussion and details, although this subject will also be explored in future studies.

Scattering from dust on the optical surfaces has been long speculated as an important contributor to the wide-angle PSF (e.g., \citealt{1996PASP..108..699R}). \cite{2002A&A...384..763M} has investigated temporal variations of the extended PSF wing. However, the measurements were separated by three months, and thus it is unclear whether/how dust plays a role in conditioning the wide-angle PSF shape. 

As has already been noted, we obtained images of the open cluster M44 on two closely separated nights: data from the first night were taken one day before the seasonal lens cleaning, when the lenses were rather dirty; data obtained on the second night were taken shortly after the lens cleaning. Both observing nights were clear, though it is not possible to control every atmospheric factor. However, by taking data on closely separated observing nights, we expect atmospheric conditions that vary on a longer timescale to be similar.

\begin{figure}[!htbp]
\centering
  \resizebox{\hsize}{!}{\includegraphics{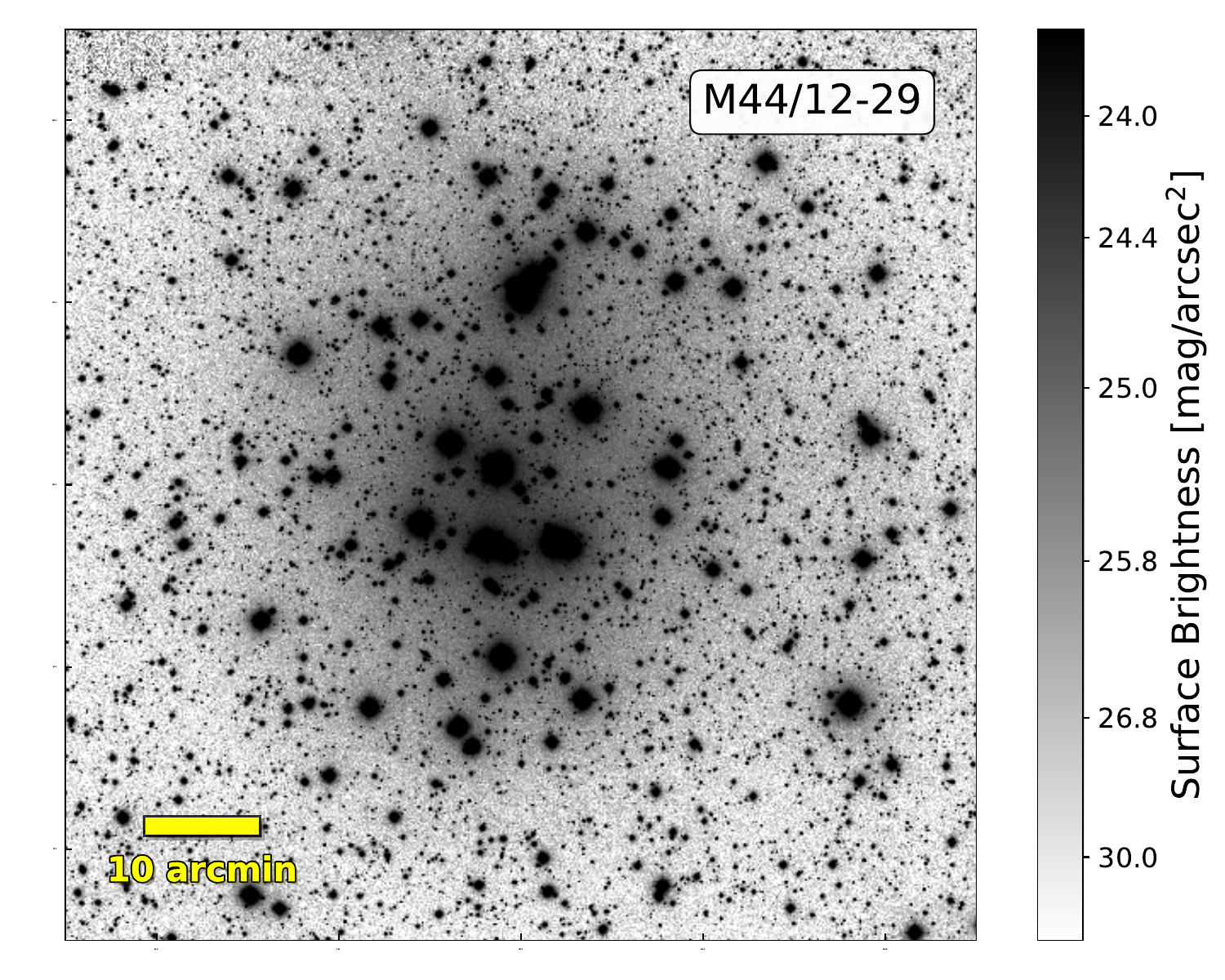}}
  \caption{1.6{\degree}$\times$1.6{\degree} g-band image of the open cluster M44 taken by Dragonfly on 2019 December 29, before lens cleaning on 2019 December 30. The image scale is stretched as in Figure \ref{fig:m44-1231_image} to enhance low surface brightness levels. The scattered light from bright stars is more prominent before lens cleaning.}
\label{fig:m44-1229_image}
\end{figure}
    
\begin{figure}[!htbp]
\centering
  \resizebox{\hsize}{!}{\includegraphics{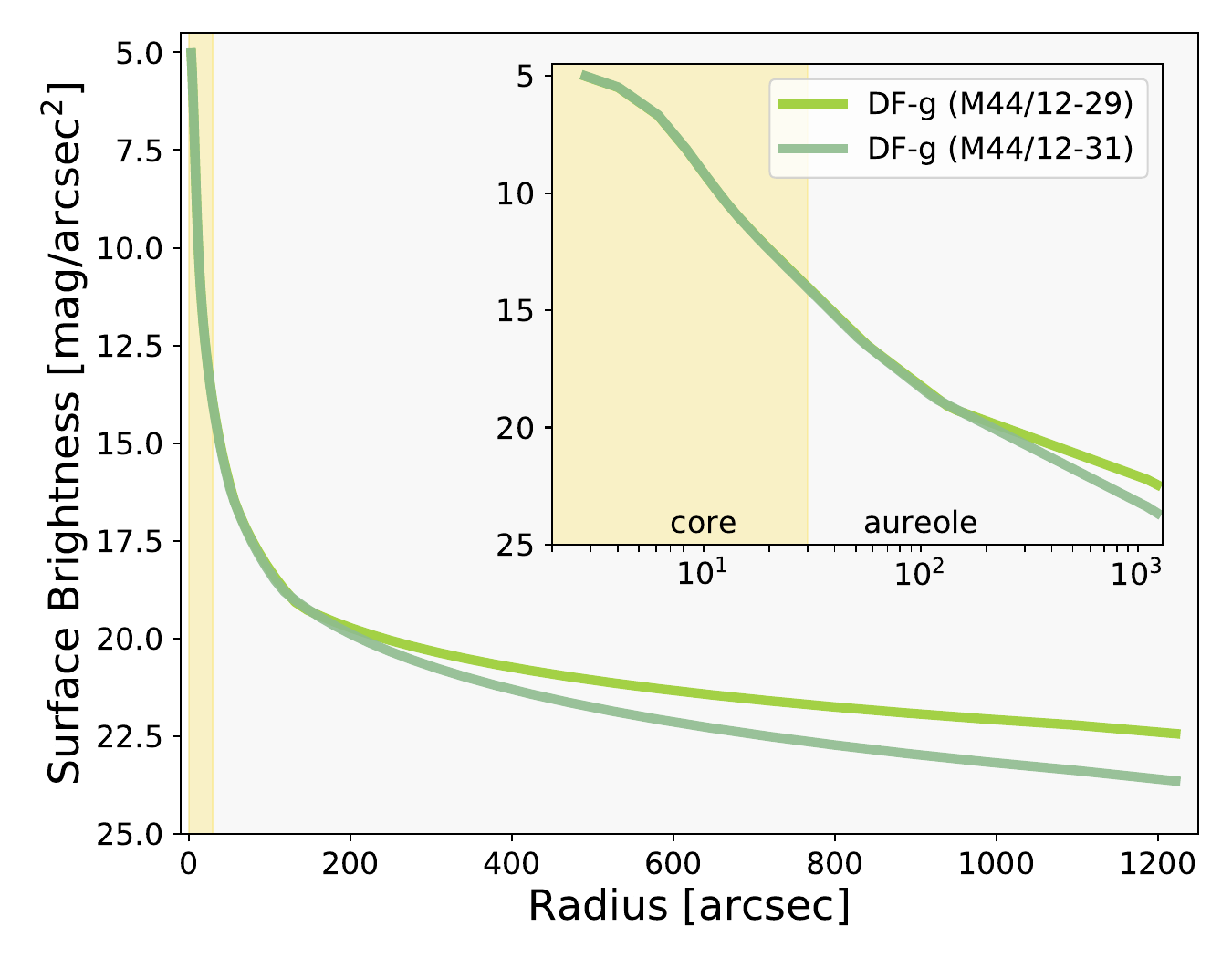}}
  \caption{Wide-angle PSF model fitted from the M44 field before (M44/12-29) and after (M44/12-31) lens cleaning on a linear (major) / log (minor) scale. The surface brightness profiles are normalized to a magnitude 0 star.
  }
\label{fig:psf_m44_2}
\end{figure}

Figure \ref{fig:m44-1229_image} displays the central cutout of the M44 image from the first night. This should be compared with the image from the second night in Figure \ref{fig:m44-1231_image}, which has been scaled identically. It is clear from visual inspection that the wide-angle PSF is more extended before lens cleaning (Figure \ref{fig:m44-1229_image}) than after lens cleaning (Figure \ref{fig:m44-1231_image}). 

In Figure \ref{fig:psf_m44_2} we show the wide-angle PSF modeled from the two nights of data. The two Dragonfly PSFs are similar out to 2$\arcmin$--3$\arcmin$, both of which follow a power law of $r^{-3.5}$ out to around 1$\arcmin$ and then flatten out to a power law of $r^{-2.8}$. Beyond 3$\arcmin$, there is noticeable difference between the PSFs obtained on the two nights. Although they both show flattening, the PSF from the first night becomes much shallower ($n=1.35$) than the one on the second night ($n=1.9$). {Therefore, our leading hypothesis is that lens cleanliness is an important element (and may be the leading element), amongst the various factors conditioning the wide-angle PSF. The relative importance of these factors may vary from night to night and will be investigated in our future studies}.

In effect, the PSF has a secular degradation as dust accumulates on the optics. For this reason, combining/analyzing low surface brightness imaging data obtained over a wide period of time should take this potential secular change in the wide-angle PSF under advisement. Careful attention to optical cleanliness is needed to achieve the optimal level of scattered light suppression for low surface brightness imaging. In any case, our approach to PSF modeling offers some prospect for correcting the secular degradation of the wide-angle PSF once it has been well modeled and if telescope internal factors dominate over atmospheric conditions, which will be explored in a future paper.

\section{Caveats} \label{Sec:Discussion}

There are several limitations in the assumptions/procedures of our PSF modeling:

(a) The subtraction of external light relies on the {assumption that bright sources are not too closely spaced}. Although this assumption works for Dragonfly in a crowded field like open clusters, the assumption that mutual distances {between} bright stars are much larger than the scale radius might break in very crowded fields, e.g., in globular clusters or near the Galactic center. Furthermore, the presence of binary stars would also affect the fitting where the normalization measured by brightness is affected by contiguous sources. Other tools implementing probabilistic optimization schemes, such as The Tractor (\citealt{2016ascl.soft04008L}), might be helpful to characterize the PSF in these cases.

(b) To minimize assumptions, we only use first-order polynomials for our smooth background model. Furthermore, noise in the background is assumed to be homogeneous across the field. However, one can envision scenarios where these assumptions would be wrong, e.g., patterns in the background from imperfect flat fielding, offsets in the background during coadding due to zero-point uncertainties, {variability in noise due to frame coverage and/or vignetting,} and other unexplained light in the background such as Galactic cirrus, etc. The Galactic cirrus, in particular, is clearly an important source of contamination, which will need to be taken into account when significant. Another possible complication is spatial variability of the wide-angle PSF. We use a fixed PSF model in a single run assuming the PSF variation is small within the region of modeling, which might not be true at large distances from the field center. While we currently restrict our modeling to scales smaller than 1\degree$\times$1\degree, in future work we will broaden our approach to encompass more comprehensive background models and variation of the wide-angle PSF.

(c) Background removal in the data reduction process remains a challenge. Most data reduction pipelines remove background structures on large scales prior to the analysis. For example, SDSS data products have a global sky subtraction on a scale of around 20{\arcmin}, with its photometric pipeline subtracting a local sky within 1{\arcmin}--2{\arcmin}. Typically, backgrounds on scales larger than the individual chip size of the detector are removed. As a consequence, large scale patterns, including the wide-angle PSF, are also partly subtracted or altered in the final output. In the Dragonfly pipeline, this issue is alleviated thanks to the very wide field-of-view of Dragonfly, where the large scale structures are preserved {up to} a scale of 45{\arcmin}. However, background subtraction would still affect the PSF modeling on scales larger than this, and thus limit the modeling range. One possible improvement to our procedures would be to avoid background subtraction completely during the reduction and include the background model in an optimized frame stacking. In this case, as noted before, the PSF variation and noise inhomogeneity, and also Galactic cirrus, will also need to be taken into account. 

(d) The reliability of the recovered PSF depends on the proximity of the forward model to the truth. In particular, we adopt a simple parameterization where the extended PSF wing is described by a multi-power law and the core by a Moffat function, ignoring all the high-order features (axis asymmetry, spikes, dips, etc.). This parameterization is based on two properties of the Dragonfly PSF: 1) its extended wing smoothly follows well-defined power laws and 2) our modeling focuses on the wide-angle part of the PSF, which is minimally affected by those effects (unlike the core of the PSF). Our approach might not work quite as well on other telescopes with more complex features in the PSFs. In any case, we will investigate a more flexible framework for non-parametric modeling in future work, blending \texttt{mrf} and \texttt{elderflower} for further improvement of PSF characterization for Dragonfly.

\section{Summary} \label{Sec:summary}

We have presented a method for characterizing the wide-angle PSF in deep wide-field imaging on a pixel-by-pixel level using Bayesian forward modeling. The method is computationally costly when compared to classical profile measurement or stacking; however, it has some advantages: (1) scattered light is incorporated as part of the model and therefore it works well in deep/crowded fields with many bright stars; (2) prior knowledge is incorporated into the modeling; and (3) our method makes the most use of the available information encoded in an image. Because we are focusing on the wide-angle part of the PSF, our current model is constructed by a combination of a fixed Moffat component in the core with a multi-power law aureole and a slowly varying sky background.

Our methodology was developed using data obtained with the Dragonfly Telephoto Array. As an example, we applied our wide-angle PSF modeling to images of the open cluster M44, where scattered light from bright stars pervades the entire field. The wide-angle PSF of Dragonfly can be well recovered out to 20{\arcmin}--25{\arcmin}. We compare the Dragonfly PSF obtained with modeling with several wide-angle PSF measurements in the literature, including SDSS, to show the power of Dragonfly in suppressing the scattered light on large scales. 
Using the two nights data of the M44 field, before and after lens cleaning, we find a temporal degradation in the wide-angle PSF {in the sense that} the PSF wing is flatter before lens cleaning. With the lens cleaned, the wide-angle PSF of Dragonfly follows a power law close to the $r^{-2}$ law found by \cite{1971PASP...83..199K}. {This suggests that dust accumulation can significantly flatten the extended wing at large radii, highlighting the importance of optical cleanliness for suppression of PSF wings for low surface brightness imaging.}

\smallskip

\section*{Acknowledgement}
Q.L. is supported by an Ontario Trillium Award. The research of R.A. and P.G.M. is supported by grants from the Natural Sciences and Engineering Research Council of Canada. J.P.G. is supported by an NSF Astronomy and Astrophysics Postdoctoral Fellowship under award AST-1801921. S.D. is supported by NASA through Hubble Fellowship grant \# HST-HF2-51454.001-A awarded by the Space Telescope Science Institute, which is operated by the Association of Universities for Research in Astronomy, Incorporated, under NASA contract NAS5-26555. The Dunlap Institute is funded through an endowment established by the David Dunlap family and the University of Toronto. The authors thank the excellent and dedicated staff at the New Mexico Skies Observatory. We also thank S{\'e}bastien Fabbro for his support and help with
the CANFAR services. Q.L. would like to thank Joshua Speagle and Howard Yee for the useful feedback and discussion. 

The Pan-STARRS1 Surveys (PS1) and the PS1 public science archive have been made possible through contributions by the Institute for Astronomy, the University of Hawaii, the Pan-STARRS Project Office, the Max-Planck Society and its participating institutes, the Max Planck Institute for Astronomy, Heidelberg and the Max Planck Institute for Extraterrestrial Physics, Garching, The Johns Hopkins University, Durham University, the University of Edinburgh, the Queen's University Belfast, the Harvard-Smithsonian Center for Astrophysics, the Las Cumbres Observatory Global Telescope Network Incorporated, the National Central University of Taiwan, the Space Telescope Science Institute, the National Aeronautics and Space Administration under Grant No. NNX08AR22G issued through the Planetary Science Division of the NASA Science Mission Directorate, the National Science Foundation Grant No. AST-1238877, the University of Maryland, Eotvos Lorand University (ELTE), the Los Alamos National Laboratory, and the Gordon and Betty Moore Foundation

\software{SExtractor \citep{1996A&AS..117..393B}, SWarp \citep{2010ascl.soft10068B}, photutils \citep{2016ascl.soft09011B}, galsim \citep{2015A&C....10..121R}, astropy \citep{astropy:2013, astropy:2018}, numpy \citep{harris2020array}, scipy \citep{2020NatMe..17..261V}, matplotlib \citep{Hunter:2007}}, dynesty \citep{2020MNRAS.493.3132S}, mrf \citep{2020PASP..132g4503V}. 



\appendix

\section{Simulation with Known PSF} \label{Sec:fit_mock}
    
    \begin{figure*}[!htbp]
    \centering
      \resizebox{0.9\hsize}{!}{\includegraphics{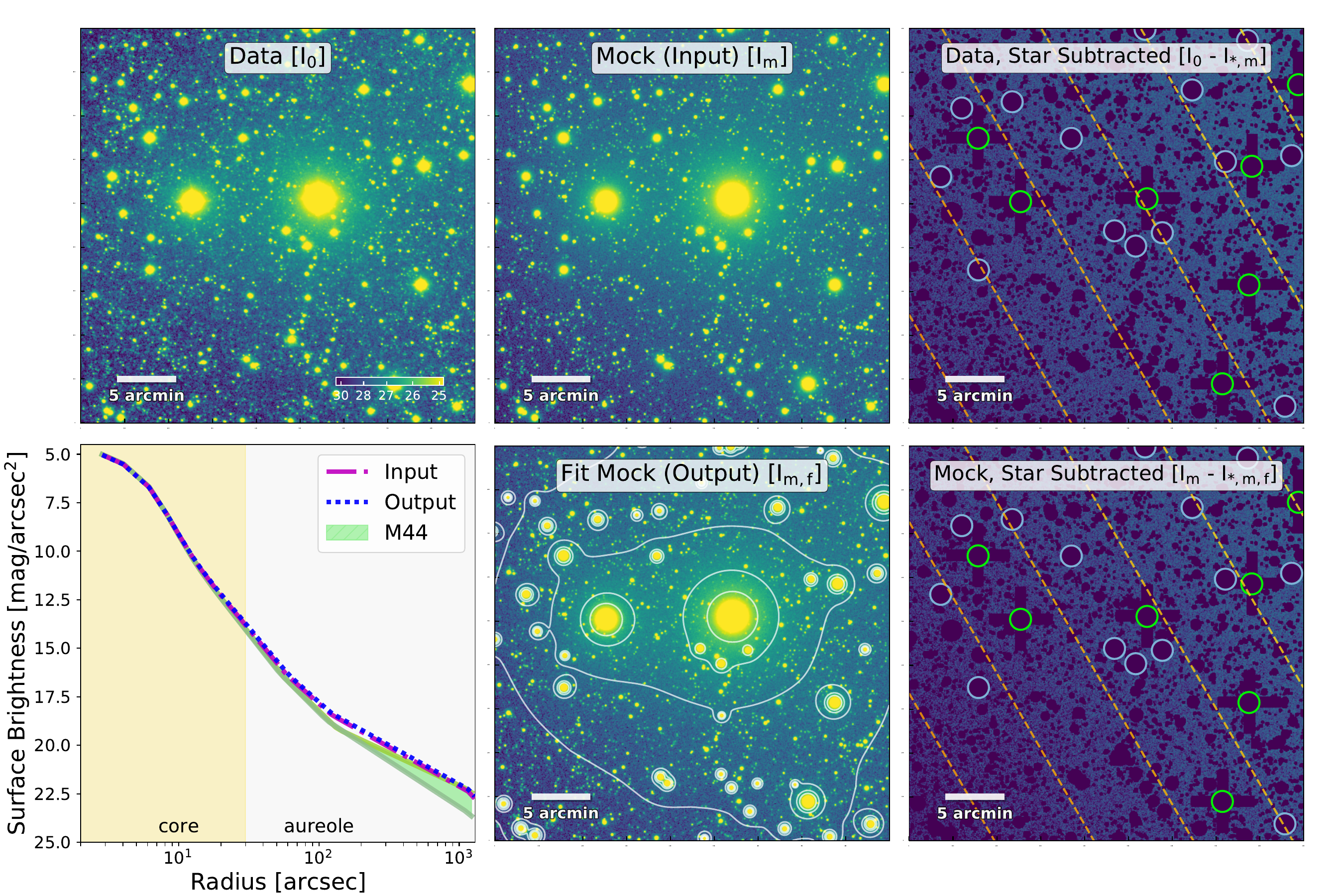}}
      \caption{{\em Top left}: a 40$\arcmin\times$40$\arcmin$ region in the Dragonfly NGC 4013 field with which the PSF model is extracted. {\em Top middle}: the simulated image with a known PSF as the input. {\em Top right}: the data after subtraction of bright stars model generated with the input PSF. {\em Bottom left}: the fitted PSF from the mock image (blue) compared with the input model (magenta). PSFs from M44 are also plotted, with the one before/after lens cleaning being the upper/lower boundary of the green band. {\em Bottom middle}: the recovered image built with the output model. Contours indicate the scattered light from bright stars (30/29/28/27/26 mag/arcsec$^2$). {\em Bottom right}: the mock image after subtraction of bright stars model generated with the output PSF. In the right panels, contours indicate the background gradient and bright stars are masked (green: VB stars; blue: MB stars). The subscript {\em$m$}/{\em$f$}/{\em$*$} stands for mock/fitted/bright star component, respectively.}
    \label{fig:fit_ngc4013_mock}
    \end{figure*}
    
    {In Section \ref{Sec:modeling_m44}, we demonstrate that our implementation is able to produce a self-consistent wide-angle PSF model by showing that the result matches the stacked stellar profile if the scattered light from other bright stars is taken into account. Another approach to testing our method is to apply it to simulated images where the PSF is known.}
    
    {To construct a simulation that closely resembles our datasets, we first extract a realistic PSF model from some representative data. We select a 40$\arcmin\times$40$\arcmin$ region in the NGC 4013 g-band image taken from the DEGS as a test field and fit the wide-angle PSF model using \texttt{elderflower}. The fitting setup is the same as the setup for the NGC 5907 field in Section \ref{Sec:ngc5907}. In addition, we include first-order polynomials in the background model. The extracted PSF model is shown as the magenta curve in the bottom left panel of Figure \ref{fig:fit_ngc4013_mock}. A mock image including all stars brighter than 22 mag is generated using \texttt{galsim} based on the extracted PSF and background model, as displayed in the top middle panel. The injected sky noise has a standard deviation of $\sigma_{\rm sky}$ from fitting. The right panel shows the residual of the data and the bright stars model (leaving out the background), demonstrating that the input model faithfully reproduces the actual data. For simplicity, we do not include extended sources in the simulated image. In practice, the bright extended sources can be masked with catalogs while faint ones would not dominate the results due to the use of a large number of pixels.}
    
    {Next, we run the model fitting on the mock image with the same setup. The resulting PSF model is shown as the blue curve in the bottom left panel of Figure \ref{fig:fit_ngc4013_mock}. The output model is in good correspondence with the input model, demonstrating the efficacy of our implementation in recovering the wide-angle PSF here. The fitted PSF parameters are consistent with the inputs within 2\%. The recovered image using the output model is displayed in the bottom middle panel. We show the residual of the input mock image and the output bright star models in the bottom right panel. The background mean is slightly higher than the input by 0.06\% ($\sim$0.07 $\sigma_{\rm sky}$). The 1$\sigma$ fractional difference of the input image and the output image (with the stellar cores masked) is within 0.5\% ($\sim$1.1 $\sigma_{\rm sky}$), comparable to the injected sky noise. Note that stars fainter than 22 mag are not rendered in the mock image, which might cause the slightly different background between the input and the output. However, as revealed by the good correspondence between the input and output, the possible perturbation on the output PSF due to background is small.}
    
    {In the bottom left panel of Figure \ref{fig:fit_ngc4013_mock}, we also plot the wide-angle PSF model obtained from the M44 field in Section \ref{Sec:modeling_m44} for comparison. The wide-angle PSF in the test field has a slightly shallower first power-law component ($r^{-3.5}$) than that of the M44 field ($r^{-3.6}$), and the power index of its outermost component ($n=1.6$) falls in between those of the M44 field before ($n=1.35$) and after ($n=1.9$) the lens cleaning.}
    
    {We have run similar simulations on mock images generated from several different regions from different fields (with various spatial and brightness distribution of stars, levels of sky noise, etc.), and the results are similar to the example test field shown in Figure \ref{fig:fit_ngc4013_mock}, with the output PSFs being consistent with the input PSFs.}

\section{Photometric Test with a Mock Galaxy} \label{Sec:mock_test}
    \begin{figure*}[!htbp]
    \centering
      \resizebox{0.9\hsize}{!}{\includegraphics{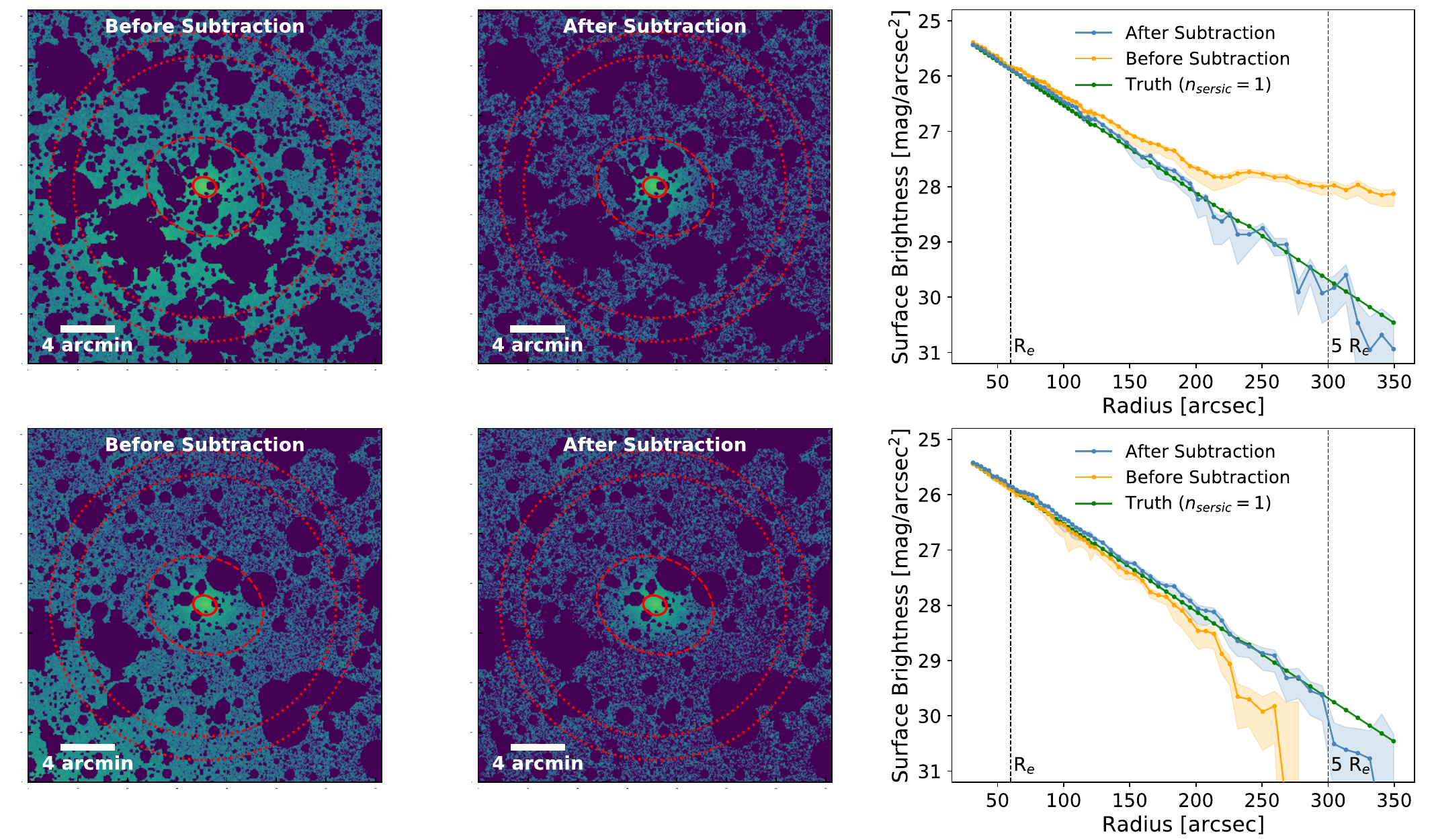}}
      \caption{Left: a mock diffuse galaxy placed at two positions in a field affected by scattered light from bright stars around NGC 5907; Middle: the same galaxy placed at the same positions in the field but with the wide-angle PSF subtracted. The range of $R_e$ and 5 $R_e$ are indicated by red solid and dashed ellipses. The ring used for measuring the background is indicated by red dotted annuli. Right: profiles extracted from the left (blue curves) and middle (orange curves) images. The model truth (the $n_{sersic}$ = 1 S{\'e}rsic profile convolved by the PSF) is shown as the green curve. The shaded areas indicate 2$\sigma$ measurement uncertainties of the profiles. 1$R_e$ and 5$R_e$ are marked as black dashed lines. Before bright-star subtraction, the profiles in the suffer from the scattered light. Subtracting the wide-angle PSF greatly improves the measurement.}
    \label{fig:test_mock}
    \end{figure*}
    
    To test whether removing the wide-angle PSF improves the photometry of galaxies imaged near bright stars, we perform a photometric test. We place a mock diffuse galaxy following a S{\'e}rsic profile in the field presented in Section \ref{Sec:ngc5907}, before and after bright star subtraction. We choose a S{\'e}rsic index of $n_{sersic}$ = 1, which corresponds to the typical S{\'e}rsic index of LSBGs in the local universe (e.g., \citealt{2015ApJ...798L..45V}, \citealt{2015ApJ...807L...2K}). The mock galaxy has a scale radius $R_e$ of 60{\arcsec}, an axis ratio of 0.8, and a position angle of 70\degree. It has a total magnitude of 14 and a central surface brightness of 25 mag/arcsec$^2$. We generate the model galaxy using \texttt{Galsim} and convolve it with the model PSF obtained by \texttt{elderflower}. The model galaxy is then placed in the two images, as displayed in the first two columns of Figure \ref{fig:test_mock}. 
    
    We then perform photometry on the two mock galaxies using elliptical annuli with the axis ratio and the position angle from  ellipse isophote fitting using \texttt{photutils}. The background value is determined from a 2{\arcmin} wide circular sky annulus of at 12 $R_e$ after 3$\sigma$ clipping. Faint stars are masked using the segmentation map generated from Pan-STARRS. Luminous stars are masked out to 2{\arcmin}. {The target galaxy is also masked during modeling. The mask of the galaxy does not affect the modeling, given its small scale compared to the field and faintness at its outskirts.} Profiles are then extracted in the range from 0.5 to 6 $R_e$ (at steps of 0.05 $R_e$ between 0.5-3 $R_e$ and 0.1 $R_e$ beyond 3 $R_e$) using the 5$\sigma$-clipped mean. The extracted profiles are shown in the third column of Figure \ref{fig:test_mock} as orange (before wide-angle PSF subtraction) and blue (after wide-angle PSF subtraction) curves. The truth (i.e., the model profile convolved with PSF) is shown as the green curve. 
    
    Before accounting for the wide-angle PSF, the extracted profiles in Figure \ref{fig:test_mock} (upper) suffer from significant flattening out of 2 $R_e$ due to the scattered light from the three bright stars, while the one in Figure \ref{fig:test_mock} (lower) shows a steepening beyond 3 $R_e$ due to an overestimated background. These correspond to the two types of bias illustrated in Section \ref{Sec:challenge}. On the other hand, the extracted profiles after star subtraction follow the truth out to 5-6 $R_e$ at surface brightness around 30 mag/arcsec$^2$. Small spikes are observed, which are probably caused by unresolved sources or high-order features of the PSF. Unresolved sources and faint stars could be removed using \texttt{mrf} to help mitigate the heavy masking. 
    
    In conclusion, this photometric test demonstrates that the scattered light from bright stars affects the profile measurement in low surface brightness imaging, but that this can be largely mitigated by modeling and subtracting scattered light from the wide-angle PSF. Note that the extracted profile of the galaxy is still as convolved by the PSF, as it would be as extracted from actual data. However, with the PSF well determined, any effects on the intrinsic galaxy profile can be assessed.

\bibliography{bibtex}{}

\begin{thebibliography}{}
\expandafter\ifx\csname natexlab\endcsname\relax\def\natexlab#1{#1}\fi
\providecommand{\url}[1]{\href{#1}{#1}}
\providecommand{\dodoi}[1]{doi:~\href{http://doi.org/#1}{\nolinkurl{#1}}}
\providecommand{\doeprint}[1]{\href{http://ascl.net/#1}{\nolinkurl{http://ascl.net/#1}}}
\providecommand{\doarXiv}[1]{\href{https://arxiv.org/abs/#1}{\nolinkurl{https://arxiv.org/abs/#1}}}

\bibitem[{{Abraham} \& {van Dokkum}(2014)}]{2014PASP..126...55A}
{Abraham}, R.~G., \& {van Dokkum}, P.~G. 2014, \pasp, 126, 55,
  \dodoi{10.1086/674875}

\bibitem[{{Aihara} {et~al.}(2011){Aihara}, {Allende Prieto}, {An}, {Anderson},
  {Aubourg}, {Balbinot}, {Beers}, {Berlind}, {Bickerton}, {Bizyaev}, {Blanton},
  {Bochanski}, {Bolton}, {Bovy}, {Brandt}, {Brinkmann}, {Brown}, {Brownstein},
  {Busca}, {Campbell}, {Carr}, {Chen}, {Chiappini}, {Comparat}, {Connolly},
  {Cortes}, {Croft}, {Cuesta}, {da Costa}, {Davenport}, {Dawson}, {Dhital},
  {Ealet}, {Ebelke}, {Edmondson}, {Eisenstein}, {Escoffier}, {Esposito},
  {Evans}, {Fan}, {Femen{\'\i}a Castell{\'a}}, {Font-Ribera}, {Frinchaboy},
  {Ge}, {Gillespie}, {Gilmore}, {Gonz{\'a}lez Hern{\'a}ndez}, {Gott}, {Gould},
  {Grebel}, {Gunn}, {Hamilton}, {Harding}, {Harris}, {Hawley}, {Hearty}, {Ho},
  {Hogg}, {Holtzman}, {Honscheid}, {Inada}, {Ivans}, {Jiang}, {Johnson},
  {Jordan}, {Jordan}, {Kazin}, {Kirkby}, {Klaene}, {Knapp}, {Kneib},
  {Kochanek}, {Koesterke}, {Kollmeier}, {Kron}, {Lampeitl}, {Lang}, {Le Goff},
  {Lee}, {Lin}, {Long}, {Loomis}, {Lucatello}, {Lundgren}, {Lupton}, {Ma},
  {MacDonald}, {Mahadevan}, {Maia}, {Makler}, {Malanushenko}, {Malanushenko},
  {Mandelbaum}, {Maraston}, {Margala}, {Masters}, {McBride}, {McGehee},
  {McGreer}, {M{\'e}nard}, {Miralda-Escud{\'e}}, {Morrison}, {Mullally},
  {Muna}, {Munn}, {Murayama}, {Myers}, {Naugle}, {Neto}, {Nguyen}, {Nichol},
  {O'Connell}, {Ogando}, {Olmstead}, {Oravetz}, {Padmanabhan},
  {Palanque-Delabrouille}, {Pan}, {Pandey}, {P{\^a}ris}, {Percival},
  {Petitjean}, {Pfaffenberger}, {Pforr}, {Phleps}, {Pichon}, {Pieri}, {Prada},
  {Price-Whelan}, {Raddick}, {Ramos}, {Reyl{\'e}}, {Rich}, {Richards}, {Rix},
  {Robin}, {Rocha-Pinto}, {Rockosi}, {Roe}, {Rollinde}, {Ross}, {Ross},
  {Rossetto}, {S{\'a}nchez}, {Sayres}, {Schlegel}, {Schlesinger}, {Schmidt},
  {Schneider}, {Sheldon}, {Shu}, {Simmerer}, {Simmons}, {Sivarani}, {Snedden},
  {Sobeck}, {Steinmetz}, {Strauss}, {Szalay}, {Tanaka}, {Thakar}, {Thomas},
  {Tinker}, {Tofflemire}, {Tojeiro}, {Tremonti}, {Vandenberg}, {Vargas
  Maga{\~n}a}, {Verde}, {Vogt}, {Wake}, {Wang}, {Weaver}, {Weinberg}, {White},
  {White}, {Yanny}, {Yasuda}, {Yeche}, \& {Zehavi}}]{2011ApJS..193...29A}
{Aihara}, H., {Allende Prieto}, C., {An}, D., {et~al.} 2011, \apjs, 193, 29,
  \dodoi{10.1088/0067-0049/193/2/29}

\bibitem[{{Astropy Collaboration} {et~al.}(2013){Astropy Collaboration},
  {Robitaille}, {Tollerud}, {Greenfield}, {Droettboom}, {Bray}, {Aldcroft},
  {Davis}, {Ginsburg}, {Price-Whelan}, {Kerzendorf}, {Conley}, {Crighton},
  {Barbary}, {Muna}, {Ferguson}, {Grollier}, {Parikh}, {Nair}, {Unther},
  {Deil}, {Woillez}, {Conseil}, {Kramer}, {Turner}, {Singer}, {Fox}, {Weaver},
  {Zabalza}, {Edwards}, {Azalee Bostroem}, {Burke}, {Casey}, {Crawford},
  {Dencheva}, {Ely}, {Jenness}, {Labrie}, {Lim}, {Pierfederici}, {Pontzen},
  {Ptak}, {Refsdal}, {Servillat}, \& {Streicher}}]{astropy:2013}
{Astropy Collaboration}, {Robitaille}, T.~P., {Tollerud}, E.~J., {et~al.} 2013,
  \aap, 558, A33, \dodoi{10.1051/0004-6361/201322068}

\bibitem[{{Astropy Collaboration} {et~al.}(2018){Astropy Collaboration},
  {Price-Whelan}, {Sip{H{o}}cz}, {G{"u}nther}, {Lim}, {Crawford}, {Conseil},
  {Shupe}, {Craig}, {Dencheva}, {Ginsburg}, {Vand erPlas}, {Bradley},
  {P{'e}rez-Su{'a}rez}, {de Val-Borro}, {Aldcroft}, {Cruz}, {Robitaille},
  {Tollerud}, {Ardelean}, {Babej}, {Bach}, {Bachetti}, {Bakanov}, {Bamford},
  {Barentsen}, {Barmby}, {Baumbach}, {Berry}, {Biscani}, {Boquien}, {Bostroem},
  {Bouma}, {Brammer}, {Bray}, {Breytenbach}, {Buddelmeijer}, {Burke},
  {Calderone}, {Cano Rodr{'i}guez}, {Cara}, {Cardoso}, {Cheedella}, {Copin},
  {Corrales}, {Crichton}, {D'Avella}, {Deil}, {Depagne}, {Dietrich}, {Donath},
  {Droettboom}, {Earl}, {Erben}, {Fabbro}, {Ferreira}, {Finethy}, {Fox},
  {Garrison}, {Gibbons}, {Goldstein}, {Gommers}, {Greco}, {Greenfield},
  {Groener}, {Grollier}, {Hagen}, {Hirst}, {Homeier}, {Horton}, {Hosseinzadeh},
  {Hu}, {Hunkeler}, {Ivezi{'c}}, {Jain}, {Jenness}, {Kanarek}, {Kendrew},
  {Kern}, {Kerzendorf}, {Khvalko}, {King}, {Kirkby}, {Kulkarni}, {Kumar},
  {Lee}, {Lenz}, {Littlefair}, {Ma}, {Macleod}, {Mastropietro}, {McCully},
  {Montagnac}, {Morris}, {Mueller}, {Mumford}, {Muna}, {Murphy}, {Nelson},
  {Nguyen}, {Ninan}, {N{"o}the}, {Ogaz}, {Oh}, {Parejko}, {Parley}, {Pascual},
  {Patil}, {Patil}, {Plunkett}, {Prochaska}, {Rastogi}, {Reddy Janga},
  {Sabater}, {Sakurikar}, {Seifert}, {Sherbert}, {Sherwood-Taylor}, {Shih},
  {Sick}, {Silbiger}, {Singanamalla}, {Singer}, {Sladen}, {Sooley},
  {Sornarajah}, {Streicher}, {Teuben}, {Thomas}, {Tremblay}, {Turner},
  {Terr{'o}n}, {van Kerkwijk}, {de la Vega}, {Watkins}, {Weaver}, {Whitmore},
  {Woillez}, {Zabalza}, \& {Astropy Contributors}}]{astropy:2018}
{Astropy Collaboration}, {Price-Whelan}, A.~M., {Sip{H{o}}cz}, B.~M., {et~al.}
  2018, aj, 156, 123, \dodoi{10.3847/1538-3881/aabc4f}

\bibitem[{{Bally} {et~al.}(2000){Bally}, {O'Dell}, \&
  {McCaughrean}}]{2000AJ....119.2919B}
{Bally}, J., {O'Dell}, C.~R., \& {McCaughrean}, M.~J. 2000, \aj, 119, 2919,
  \dodoi{10.1086/301385}

\bibitem[{{Beckers}(1995)}]{1995seft.conf..303B}
{Beckers}, J.~M. 1995, in Scientific and Engineering Frontiers for 8 - 10 m
  Telescopes, 303--312

\bibitem[{{Berg{\'e}} {et~al.}(2012){Berg{\'e}}, {Price}, {Amara}, \&
  {Rhodes}}]{2012MNRAS.419.2356B}
{Berg{\'e}}, J., {Price}, S., {Amara}, A., \& {Rhodes}, J. 2012, \mnras, 419,
  2356, \dodoi{10.1111/j.1365-2966.2011.19888.x}

\bibitem[{{Bernstein}(2007)}]{2007ApJ...666..663B}
{Bernstein}, R.~A. 2007, \apj, 666, 663, \dodoi{10.1086/519824}

\bibitem[{{Bertin}(2010)}]{2010ascl.soft10068B}
{Bertin}, E. 2010, {SWarp: Resampling and Co-adding FITS Images Together}.
\newblock \doeprint{1010.068}

\bibitem[{{Bertin}(2013)}]{2013ascl.soft01001B}
---. 2013, {PSFEx: Point-Spread Function Extractor}.
\newblock \doeprint{1301.001}

\bibitem[{{Bertin} \& {Arnouts}(1996)}]{1996A&AS..117..393B}
{Bertin}, E., \& {Arnouts}, S. 1996, \aaps, 117, 393,
  \dodoi{10.1051/aas:1996164}

\bibitem[{{Blanton} {et~al.}(2011){Blanton}, {Kazin}, {Muna}, {Weaver}, \&
  {Price-Whelan}}]{2011AJ....142...31B}
{Blanton}, M.~R., {Kazin}, E., {Muna}, D., {Weaver}, B.~A., \& {Price-Whelan},
  A. 2011, \aj, 142, 31, \dodoi{10.1088/0004-6256/142/1/31}

\bibitem[{{Bradley} {et~al.}(2016){Bradley}, {Sipocz}, {Robitaille},
  {Tollerud}, {Deil}, {Vin{\'\i}cius}, {Barbary}, {G{\"u}nther}, {Bostroem},
  {Droettboom}, {Bray}, {Bratholm}, {Pickering}, {Craig}, {Pascual}, {Greco},
  {Donath}, {Kerzendorf}, {Littlefair}, {Barentsen}, {D'Eugenio}, \&
  {Weaver}}]{2016ascl.soft09011B}
{Bradley}, L., {Sipocz}, B., {Robitaille}, T., {et~al.} 2016, {Photutils:
  Photometry Tools}.
\newblock \doeprint{1609.011}

\bibitem[{{Brainerd} {et~al.}(1996){Brainerd}, {Blandford}, \&
  {Smail}}]{1996ApJ...466..623B}
{Brainerd}, T.~G., {Blandford}, R.~D., \& {Smail}, I. 1996, \apj, 466, 623,
  \dodoi{10.1086/177537}

\bibitem[{{Capaccioli} \& {de Vaucouleurs}(1983)}]{1983ApJS...52..465C}
{Capaccioli}, M., \& {de Vaucouleurs}, G. 1983, \apjs, 52, 465,
  \dodoi{10.1086/190879}

\bibitem[{{Chambers} {et~al.}(2016){Chambers}, {Magnier}, {Metcalfe},
  {Flewelling}, {Huber}, {Waters}, {Denneau}, {Draper}, {Farrow}, {Finkbeiner},
  {Holmberg}, {Koppenhoefer}, {Price}, {Rest}, {Saglia}, {Schlafly}, {Smartt},
  {Sweeney}, {Wainscoat}, {Burgett}, {Chastel}, {Grav}, {Heasley}, {Hodapp},
  {Jedicke}, {Kaiser}, {Kudritzki}, {Luppino}, {Lupton}, {Monet}, {Morgan},
  {Onaka}, {Shiao}, {Stubbs}, {Tonry}, {White}, {Ba{\~n}ados}, {Bell},
  {Bender}, {Bernard}, {Boegner}, {Boffi}, {Botticella}, {Calamida},
  {Casertano}, {Chen}, {Chen}, {Cole}, {Deacon}, {Frenk}, {Fitzsimmons},
  {Gezari}, {Gibbs}, {Goessl}, {Goggia}, {Gourgue}, {Goldman}, {Grant},
  {Grebel}, {Hambly}, {Hasinger}, {Heavens}, {Heckman}, {Henderson}, {Henning},
  {Holman}, {Hopp}, {Ip}, {Isani}, {Jackson}, {Keyes}, {Koekemoer}, {Kotak},
  {Le}, {Liska}, {Long}, {Lucey}, {Liu}, {Martin}, {Masci}, {McLean}, {Mindel},
  {Misra}, {Morganson}, {Murphy}, {Obaika}, {Narayan}, {Nieto-Santisteban},
  {Norberg}, {Peacock}, {Pier}, {Postman}, {Primak}, {Rae}, {Rai}, {Riess},
  {Riffeser}, {Rix}, {R{\"o}ser}, {Russel}, {Rutz}, {Schilbach}, {Schultz},
  {Scolnic}, {Strolger}, {Szalay}, {Seitz}, {Small}, {Smith}, {Soderblom},
  {Taylor}, {Thomson}, {Taylor}, {Thakar}, {Thiel}, {Thilker}, {Unger},
  {Urata}, {Valenti}, {Wagner}, {Walder}, {Walter}, {Watters}, {Werner},
  {Wood-Vasey}, \& {Wyse}}]{2016arXiv161205560C}
{Chambers}, K.~C., {Magnier}, E.~A., {Metcalfe}, N., {et~al.} 2016, arXiv
  e-prints, arXiv:1612.05560.
\newblock \doarXiv{1612.05560}

\bibitem[{{Coupon} {et~al.}(2018){Coupon}, {Czakon}, {Bosch}, {Komiyama},
  {Medezinski}, {Miyazaki}, \& {Oguri}}]{2018PASJ...70S...7C}
{Coupon}, J., {Czakon}, N., {Bosch}, J., {et~al.} 2018, \pasj, 70, S7,
  \dodoi{10.1093/pasj/psx047}

\bibitem[{{Danieli} {et~al.}(2020){Danieli}, {Lokhorst}, {Zhang}, {Merritt},
  {van Dokkum}, {Abraham}, {Conroy}, {Gilhuly}, {Greco}, {Janssens}, {Li},
  {Liu}, {Miller}, \& {Mowla}}]{2020ApJ...894..119D}
{Danieli}, S., {Lokhorst}, D., {Zhang}, J., {et~al.} 2020, \apj, 894, 119,
  \dodoi{10.3847/1538-4357/ab88a8}

\bibitem[{{de Jong}(2008)}]{2008MNRAS.388.1521D}
{de Jong}, R.~S. 2008, \mnras, 388, 1521,
  \dodoi{10.1111/j.1365-2966.2008.13505.x}

\bibitem[{{DeVore} {et~al.}(2013){DeVore}, {Kristl}, \&
  {Rappaport}}]{2013JGRD..118.5679D}
{DeVore}, J.~G., {Kristl}, J.~A., \& {Rappaport}, S.~A. 2013, Journal of
  Geophysical Research (Atmospheres), 118, 5679, \dodoi{10.1002/jgrd.50440}

\bibitem[{{D'Souza} {et~al.}(2014){D'Souza}, {Kauffman}, {Wang}, \&
  {Vegetti}}]{2014MNRAS.443.1433D}
{D'Souza}, R., {Kauffman}, G., {Wang}, J., \& {Vegetti}, S. 2014, \mnras, 443,
  1433, \dodoi{10.1093/mnras/stu1194}

\bibitem[{{F{\'e}tick} {et~al.}(2020){F{\'e}tick}, {Mugnier}, {Fusco}, \&
  {Neichel}}]{2020MNRAS.496.4209F}
{F{\'e}tick}, R.~J.~L., {Mugnier}, L.~M., {Fusco}, T., \& {Neichel}, B. 2020,
  \mnras, 496, 4209, \dodoi{10.1093/mnras/staa1813}

\bibitem[{{Fischer} {et~al.}(2017){Fischer}, {Bernardi}, \&
  {Meert}}]{2017MNRAS.467..490F}
{Fischer}, J.~L., {Bernardi}, M., \& {Meert}, A. 2017, \mnras, 467, 490,
  \dodoi{10.1093/mnras/stx136}

\bibitem[{{Gilhuly} {et~al.}(2020){Gilhuly}, {Hendel}, {Merritt}, {Abraham},
  {Danieli}, {Lokhorst}, {Liu}, {van Dokkum}, {Conroy}, \&
  {Greco}}]{2020ApJ...897..108G}
{Gilhuly}, C., {Hendel}, D., {Merritt}, A., {et~al.} 2020, \apj, 897, 108,
  \dodoi{10.3847/1538-4357/ab9b25}

\bibitem[{{Guhathakurta} \& {Tyson}(1989)}]{1989ApJ...346..773G}
{Guhathakurta}, P., \& {Tyson}, J.~A. 1989, \apj, 346, 773,
  \dodoi{10.1086/168058}

\bibitem[{Harris {et~al.}(2020)Harris, Millman, van~der Walt, Gommers,
  Virtanen, Cournapeau, Wieser, Taylor, Berg, Smith, Kern, Picus, Hoyer, van
  Kerkwijk, Brett, Haldane, del R{'{\i}}o, Wiebe, Peterson,
  G{'{e}}rard-Marchant, Sheppard, Reddy, Weckesser, Abbasi, Gohlke, \&
  Oliphant}]{harris2020array}
Harris, C.~R., Millman, K.~J., van~der Walt, S.~J., {et~al.} 2020, Nature, 585,
  357, \dodoi{10.1038/s41586-020-2649-2}

\bibitem[{{Herbel} {et~al.}(2018){Herbel}, {Kacprzak}, {Amara}, {Refregier}, \&
  {Lucchi}}]{2018JCAP...07..054H}
{Herbel}, J., {Kacprzak}, T., {Amara}, A., {Refregier}, A., \& {Lucchi}, A.
  2018, \jcap, 2018, 054, \dodoi{10.1088/1475-7516/2018/07/054}

\bibitem[{{Huang} {et~al.}(2018){Huang}, {Leauthaud}, {Greene}, {Bundy}, {Lin},
  {Tanaka}, {Miyazaki}, \& {Komiyama}}]{2018MNRAS.475.3348H}
{Huang}, S., {Leauthaud}, A., {Greene}, J.~E., {et~al.} 2018, \mnras, 475,
  3348, \dodoi{10.1093/mnras/stx3200}

\bibitem[{Hunter(2007)}]{Hunter:2007}
Hunter, J.~D. 2007, Computing in Science \& Engineering, 9, 90,
  \dodoi{10.1109/MCSE.2007.55}

\bibitem[{{Infante-Sainz} {et~al.}(2020){Infante-Sainz}, {Trujillo}, \&
  {Rom{\'a}n}}]{2020MNRAS.491.5317I}
{Infante-Sainz}, R., {Trujillo}, I., \& {Rom{\'a}n}, J. 2020, \mnras, 491,
  5317, \dodoi{10.1093/mnras/stz3111}

\bibitem[{{Jarvis} {et~al.}(2021){Jarvis}, {Bernstein}, {Amon}, {Davis},
  {L{\'e}get}, {Bechtol}, {Harrison}, {Gatti}, {Roodman}, {Chang}, {Chen},
  {Choi}, {Desai}, {Drlica-Wagner}, {Gruen}, {Gruendl}, {Hernandez},
  {MacCrann}, {Meyers}, {Navarro-Alsina}, {Pandey}, {Plazas}, {Secco},
  {Sheldon}, {Troxel}, {Vorperian}, {Wei}, {Zuntz}, {Abbott}, {Aguena},
  {Allam}, {Avila}, {Bhargava}, {Bridle}, {Brooks}, {Carnero Rosell}, {Carrasco
  Kind}, {Carretero}, {Costanzi}, {da Costa}, {De Vicente}, {Diehl}, {Doel},
  {Everett}, {Flaugher}, {Fosalba}, {Frieman}, {Garc{\'\i}a-Bellido},
  {Gaztanaga}, {Gerdes}, {Gutierrez}, {Hinton}, {Hollowood}, {Honscheid},
  {James}, {Kent}, {Kuehn}, {Kuropatkin}, {Lahav}, {Maia}, {March}, {Marshall},
  {Melchior}, {Menanteau}, {Miquel}, {Ogando}, {Paz-Chinch{\'o}n}, {Rykoff},
  {Sanchez}, {Scarpine}, {Schubnell}, {Serrano}, {Sevilla-Noarbe}, {Smith},
  {Suchyta}, {Swanson}, {Tarle}, {Varga}, {Walker}, {Wester}, {Wilkinson}, \&
  {(DES Collaboration)}}]{2021MNRAS.501.1282J}
{Jarvis}, M., {Bernstein}, G.~M., {Amon}, A., {et~al.} 2021, \mnras, 501, 1282,
  \dodoi{10.1093/mnras/staa3679}

\bibitem[{{Jee} {et~al.}(2007){Jee}, {Blakeslee}, {Sirianni}, {Martel},
  {White}, \& {Ford}}]{2007PASP..119.1403J}
{Jee}, M.~J., {Blakeslee}, J.~P., {Sirianni}, M., {et~al.} 2007, \pasp, 119,
  1403, \dodoi{10.1086/524849}

\bibitem[{{Ji} {et~al.}(2018){Ji}, {Hasan}, {Schmidt}, \&
  {Tyson}}]{2018PASP..130h4504J}
{Ji}, I., {Hasan}, I., {Schmidt}, S.~J., \& {Tyson}, J.~A. 2018, \pasp, 130,
  084504, \dodoi{10.1088/1538-3873/aac4ed}

\bibitem[{{Karabal} {et~al.}(2017){Karabal}, {Duc}, {Kuntschner}, {Chanial},
  {Cuillandre}, \& {Gwyn}}]{2017A&A...601A..86K}
{Karabal}, E., {Duc}, P.~A., {Kuntschner}, H., {et~al.} 2017, \aap, 601, A86,
  \dodoi{10.1051/0004-6361/201629974}

\bibitem[{{King}(1971)}]{1971PASP...83..199K}
{King}, I.~R. 1971, \pasp, 83, 199, \dodoi{10.1086/129100}

\bibitem[{{Koda} {et~al.}(2015){Koda}, {Yagi}, {Yamanoi}, \&
  {Komiyama}}]{2015ApJ...807L...2K}
{Koda}, J., {Yagi}, M., {Yamanoi}, H., \& {Komiyama}, Y. 2015, \apjl, 807, L2,
  \dodoi{10.1088/2041-8205/807/1/L2}

\bibitem[{{Kolmogorov}(1941)}]{1941DoSSR..30..301K}
{Kolmogorov}, A. 1941, Akademiia Nauk SSSR Doklady, 30, 301

\bibitem[{{Kormendy}(1973)}]{1973PASP...85..533K}
{Kormendy}, J. 1973, \pasp, 85, 533

\bibitem[{{Kraus} \& {Hillenbrand}(2007)}]{2007AJ....134.2340K}
{Kraus}, A.~L., \& {Hillenbrand}, L.~A. 2007, \aj, 134, 2340,
  \dodoi{10.1086/522831}

\bibitem[{{Lang} {et~al.}(2016){Lang}, {Hogg}, \&
  {Mykytyn}}]{2016ascl.soft04008L}
{Lang}, D., {Hogg}, D.~W., \& {Mykytyn}, D. 2016, {The Tractor: Probabilistic
  astronomical source detection and measurement}.
\newblock \doeprint{1604.008}

\bibitem[{Liu(2021)}]{elderflower}
Liu, Q. 2021, {elderflower: Wide-angle PSF Modeling in Deep Wide Images}, v0.2,
   Zenodo, \dodoi{10.5281/zenodo.5318988}

\bibitem[{{Magain} {et~al.}(2007){Magain}, {Courbin}, {Gillon}, {Sohy},
  {Letawe}, {Chantry}, \& {Letawe}}]{2007A&A...461..373M}
{Magain}, P., {Courbin}, F., {Gillon}, M., {et~al.} 2007, \aap, 461, 373,
  \dodoi{10.1051/0004-6361:20042505}

\bibitem[{{Magnier} {et~al.}(2020){Magnier}, {Schlafly}, {Finkbeiner}, {Tonry},
  {Goldman}, {R{\"o}ser}, {Schilbach}, {Casertano}, {Chambers}, {Flewelling},
  {Huber}, {Price}, {Sweeney}, {Waters}, {Denneau}, {Draper}, {Hodapp},
  {Jedicke}, {Kaiser}, {Kudritzki}, {Metcalfe}, {Stubbs}, \&
  {Wainscoat}}]{2020ApJS..251....6M}
{Magnier}, E.~A., {Schlafly}, E.~F., {Finkbeiner}, D.~P., {et~al.} 2020, \apjs,
  251, 6, \dodoi{10.3847/1538-4365/abb82a}

\bibitem[{{Mandelbaum} {et~al.}(2018){Mandelbaum}, {Lanusse}, {Leauthaud},
  {Armstrong}, {Simet}, {Miyatake}, {Meyers}, {Bosch}, {Murata}, {Miyazaki}, \&
  {Tanaka}}]{2018MNRAS.481.3170M}
{Mandelbaum}, R., {Lanusse}, F., {Leauthaud}, A., {et~al.} 2018, \mnras, 481,
  3170, \dodoi{10.1093/mnras/sty2420}

\bibitem[{{Mart{\'\i}nez-Delgado} {et~al.}(2010){Mart{\'\i}nez-Delgado},
  {Gabany}, {Crawford}, {Zibetti}, {Majewski}, {Rix}, {Fliri},
  {Carballo-Bello}, {Bardalez-Gagliuffi}, {Pe{\~n}arrubia}, {Chonis}, {Madore},
  {Trujillo}, {Schirmer}, \& {McDavid}}]{2010AJ....140..962M}
{Mart{\'\i}nez-Delgado}, D., {Gabany}, R.~J., {Crawford}, K., {et~al.} 2010,
  \aj, 140, 962, \dodoi{10.1088/0004-6256/140/4/962}

\bibitem[{{Mart{\'\i}nez-Lombilla} {et~al.}(2019){Mart{\'\i}nez-Lombilla},
  {Trujillo}, \& {Knapen}}]{2019MNRAS.483..664M}
{Mart{\'\i}nez-Lombilla}, C., {Trujillo}, I., \& {Knapen}, J.~H. 2019, \mnras,
  483, 664, \dodoi{10.1093/mnras/sty2886}

\bibitem[{{Merritt} {et~al.}(2020){Merritt}, {Pillepich}, {van Dokkum},
  {Nelson}, {Hernquist}, {Marinacci}, \& {Vogelsberger}}]{2020MNRAS.495.4570M}
{Merritt}, A., {Pillepich}, A., {van Dokkum}, P., {et~al.} 2020, \mnras, 495,
  4570, \dodoi{10.1093/mnras/staa1164}

\bibitem[{{Michard}(2002)}]{2002A&A...384..763M}
{Michard}, R. 2002, \aap, 384, 763, \dodoi{10.1051/0004-6361:20011813}

\bibitem[{{Mihos} {et~al.}(2005){Mihos}, {Harding}, {Feldmeier}, \&
  {Morrison}}]{2005ApJ...631L..41M}
{Mihos}, J.~C., {Harding}, P., {Feldmeier}, J., \& {Morrison}, H. 2005, \apjl,
  631, L41, \dodoi{10.1086/497030}

\bibitem[{{Miller} {et~al.}(2021){Miller}, {van Dokkum}, {Danieli}, {Li},
  {Abraham}, {Conroy}, {Gilhuly}, {Greco}, {Liu}, {Lokhorst}, \&
  {Merritt}}]{2021ApJ...909...74M}
{Miller}, T.~B., {van Dokkum}, P., {Danieli}, S., {et~al.} 2021, \apj, 909, 74,
  \dodoi{10.3847/1538-4357/abd7f8}

\bibitem[{{Moffat}(1969)}]{1969A&A.....3..455M}
{Moffat}, A.~F.~J. 1969, \aap, 3, 455

\bibitem[{{Morrison} {et~al.}(1994){Morrison}, {Boroson}, \&
  {Harding}}]{1994AJ....108.1191M}
{Morrison}, H.~L., {Boroson}, T.~A., \& {Harding}, P. 1994, \aj, 108, 1191,
  \dodoi{10.1086/117148}

\bibitem[{{Nelson} {et~al.}(2008){Nelson}, {Tomczyk}, {Elmore}, \&
  {Kolinski}}]{2008SPIE.7012E..31N}
{Nelson}, P.~G., {Tomczyk}, S., {Elmore}, D.~F., \& {Kolinski}, D.~J. 2008, in
  Society of Photo-Optical Instrumentation Engineers (SPIE) Conference Series,
  Vol. 7012, Ground-based and Airborne Telescopes II, ed. L.~M. {Stepp} \&
  R.~{Gilmozzi}, 701231, \dodoi{10.1117/12.789494}

\bibitem[{{Racine}(1996)}]{1996PASP..108..699R}
{Racine}, R. 1996, \pasp, 108, 699, \dodoi{10.1086/133788}

\bibitem[{{Rowe} {et~al.}(2015){Rowe}, {Jarvis}, {Mandelbaum}, {Bernstein},
  {Bosch}, {Simet}, {Meyers}, {Kacprzak}, {Nakajima}, {Zuntz}, {Miyatake},
  {Dietrich}, {Armstrong}, {Melchior}, \& {Gill}}]{2015A&C....10..121R}
{Rowe}, B.~T.~P., {Jarvis}, M., {Mandelbaum}, R., {et~al.} 2015, Astronomy and
  Computing, 10, 121, \dodoi{10.1016/j.ascom.2015.02.002}

\bibitem[{{Sandin}(2014)}]{2014A&A...567A..97S}
{Sandin}, C. 2014, \aap, 567, A97, \dodoi{10.1051/0004-6361/201423429}

\bibitem[{{Sharma}(2017)}]{2017ARA&A..55..213S}
{Sharma}, S. 2017, \araa, 55, 213, \dodoi{10.1146/annurev-astro-082214-122339}

\bibitem[{{Shectman}(1974)}]{1974ApJ...188..233S}
{Shectman}, S.~A. 1974, \apj, 188, 233, \dodoi{10.1086/152710}

\bibitem[{{Sirianni} {et~al.}(1998){Sirianni}, {Clampin}, {Hartig}, {Rafal},
  {Ford}, {Golimowski}, {Tremonti}, {Illingworth}, {Blouke}, {Lesser},
  {Burmester}, {Kimble}, {Sullivan}, {Krebs}, \&
  {Yagelowicz}}]{1998SPIE.3355..608S}
{Sirianni}, M., {Clampin}, M., {Hartig}, G.~F., {et~al.} 1998, in Society of
  Photo-Optical Instrumentation Engineers (SPIE) Conference Series, Vol. 3355,
  Optical Astronomical Instrumentation, ed. S.~{D'Odorico}, 608--612,
  \dodoi{10.1117/12.316832}

\bibitem[{{Skilling}(2004)}]{2004AIPC..735..395S}
{Skilling}, J. 2004, in American Institute of Physics Conference Series, Vol.
  735, Bayesian Inference and Maximum Entropy Methods in Science and
  Engineering: 24th International Workshop on Bayesian Inference and Maximum
  Entropy Methods in Science and Engineering, ed. R.~{Fischer}, R.~{Preuss}, \&
  U.~V. {Toussaint}, 395--405, \dodoi{10.1063/1.1835238}

\bibitem[{{Slater} {et~al.}(2009){Slater}, {Harding}, \&
  {Mihos}}]{2009PASP..121.1267S}
{Slater}, C.~T., {Harding}, P., \& {Mihos}, J.~C. 2009, \pasp, 121, 1267,
  \dodoi{10.1086/648457}

\bibitem[{{Speagle}(2020)}]{2020MNRAS.493.3132S}
{Speagle}, J.~S. 2020, \mnras, 493, 3132, \dodoi{10.1093/mnras/staa278}

\bibitem[{Stanier {et~al.}(2004)Stanier, Khlystov, \& Pandis}]{STANIER20043275}
Stanier, C.~O., Khlystov, A.~Y., \& Pandis, S.~N. 2004, Atmospheric
  Environment, 38, 3275, \dodoi{https://doi.org/10.1016/j.atmosenv.2004.03.020}

\bibitem[{{Stetson}(1987)}]{1987PASP...99..191S}
{Stetson}, P.~B. 1987, \pasp, 99, 191, \dodoi{10.1086/131977}

\bibitem[{{Tal} \& {van Dokkum}(2011)}]{2011ApJ...731...89T}
{Tal}, T., \& {van Dokkum}, P.~G. 2011, \apj, 731, 89,
  \dodoi{10.1088/0004-637X/731/2/89}

\bibitem[{{Trujillo} \& {Fliri}(2016)}]{2016ApJ...823..123T}
{Trujillo}, I., \& {Fliri}, J. 2016, \apj, 823, 123,
  \dodoi{10.3847/0004-637X/823/2/123}

\bibitem[{{van Dokkum} {et~al.}(2019){van Dokkum}, {Gilhuly}, {Bonaca},
  {Merritt}, {Danieli}, {Lokhorst}, {Abraham}, {Conroy}, \&
  {Greco}}]{2019ApJ...883L..32V}
{van Dokkum}, P., {Gilhuly}, C., {Bonaca}, A., {et~al.} 2019, \apjl, 883, L32,
  \dodoi{10.3847/2041-8213/ab40c9}

\bibitem[{{van Dokkum} {et~al.}(2020){van Dokkum}, {Lokhorst}, {Danieli}, {Li},
  {Merritt}, {Abraham}, {Gilhuly}, {Greco}, \& {Liu}}]{2020PASP..132g4503V}
{van Dokkum}, P., {Lokhorst}, D., {Danieli}, S., {et~al.} 2020, \pasp, 132,
  074503, \dodoi{10.1088/1538-3873/ab9416}

\bibitem[{{van Dokkum} {et~al.}(2015){van Dokkum}, {Abraham}, {Merritt},
  {Zhang}, {Geha}, \& {Conroy}}]{2015ApJ...798L..45V}
{van Dokkum}, P.~G., {Abraham}, R., {Merritt}, A., {et~al.} 2015, \apjl, 798,
  L45, \dodoi{10.1088/2041-8205/798/2/L45}

\bibitem[{{Virtanen} {et~al.}(2020){Virtanen}, {Gommers}, {Oliphant},
  {Haberland}, {Reddy}, {Cournapeau}, {Burovski}, {Peterson}, {Weckesser},
  {Bright}, {van der Walt}, {Brett}, {Wilson}, {Millman}, {Mayorov}, {Nelson},
  {Jones}, {Kern}, {Larson}, {Carey}, {Polat}, {Feng}, {Moore}, {VanderPlas},
  {Laxalde}, {Perktold}, {Cimrman}, {Henriksen}, {Quintero}, {Harris},
  {Archibald}, {Ribeiro}, {Pedregosa}, {van Mulbregt}, \& {SciPy 1. 0
  Contributors}}]{2020NatMe..17..261V}
{Virtanen}, P., {Gommers}, R., {Oliphant}, T.~E., {et~al.} 2020, Nature
  Methods, 17, 261, \dodoi{10.1038/s41592-019-0686-2}

\bibitem[{{Watkins} {et~al.}(2019){Watkins}, {Laine}, {Comer{\'o}n}, {Janz}, \&
  {Salo}}]{2019A&A...625A..36W}
{Watkins}, A.~E., {Laine}, J., {Comer{\'o}n}, S., {Janz}, J., \& {Salo}, H.
  2019, \aap, 625, A36, \dodoi{10.1051/0004-6361/201935130}

\bibitem[{{Watkins} {et~al.}(2016){Watkins}, {Mihos}, \&
  {Harding}}]{Watkins2016}
{Watkins}, A.~E., {Mihos}, J.~C., \& {Harding}, P. 2016, \apj, 826, 59,
  \dodoi{10.3847/0004-637X/826/1/59}

\bibitem[{{Zhang} {et~al.}(2019){Zhang}, {Yanny}, {Palmese}, {Gruen}, {To},
  {Rykoff}, {Leung}, {Collins}, {Hilton}, {Abbott}, {Annis}, {Avila}, {Bertin},
  {Brooks}, {Burke}, {Carnero Rosell}, {Carrasco Kind}, {Carretero}, {Cunha},
  {D'Andrea}, {da Costa}, {De Vicente}, {Desai}, {Diehl}, {Dietrich}, {Doel},
  {Drlica-Wagner}, {Eifler}, {Evrard}, {Flaugher}, {Fosalba}, {Frieman},
  {Garc{\'\i}a-Bellido}, {Gaztanaga}, {Gerdes}, {Gruendl}, {Gschwend},
  {Gutierrez}, {Hartley}, {Hollowood}, {Honscheid}, {Hoyle}, {James},
  {Jeltema}, {Kuehn}, {Kuropatkin}, {Li}, {Lima}, {Maia}, {March}, {Marshall},
  {Melchior}, {Menanteau}, {Miller}, {Miquel}, {Mohr}, {Ogando}, {Plazas},
  {Romer}, {Sanchez}, {Scarpine}, {Schubnell}, {Serrano}, {Sevilla-Noarbe},
  {Smith}, {Soares-Santos}, {Sobreira}, {Suchyta}, {Swanson}, {Tarle},
  {Thomas}, {Wester}, \& {DES Collaboration}}]{2019ApJ...874..165Z}
{Zhang}, Y., {Yanny}, B., {Palmese}, A., {et~al.} 2019, \apj, 874, 165,
  \dodoi{10.3847/1538-4357/ab0dfd}

\end{thebibliography}
\bibliographystyle{aasjournal}



\end{document}